\documentclass[useAMS,usenatbib]{mn2e}
\usepackage{graphicx}

\title[Photometric Properties of Void Galaxies in the
SDSS DR7 Date Release]{Photometric Properties of Void Galaxies in the Sloan Digital Sky
  Survey DR7 Data Release\\ } 
\author[F. Hoyle et al.]
  { Fiona~Hoyle,$^{1*}$, M.S.~Vogeley,$^2$ \&  D.~Pan,$^{2,3}$ \\
  $^1$ Pontifica Universidad Catolica de Ecuador, 12 de Octubre 1076 y Roca,
  Quito, Ecuador \\
  $^2$Drexel University, 3141 Chestnut Street, Philadelphia PA 19104,
  USA \\
   $^{2, 3}$ Shanghai Astronomical Observatory, Shanghai, China, 200030 \\
$^*$ fionahoyle11@gmail.com }
\date{Released 2012 Xxxxx XX}

\pagerange{\pageref{firstpage}--\pageref{lastpage}} \pubyear{2012}

\def\LaTeX{L\kern-.36em\raise.3ex\hbox{a}\kern-.15em
    T\kern-.1667em\lower.7ex\hbox{E}\kern-.125emX}

\begin{document}

\label{firstpage}

\maketitle

\begin{abstract}
Using the sample presented in \citet{Pan:2011}, we analyse the
photometric properties of 88,794 void galaxies and compare them to
galaxies that reside in higher density environments with the same
absolute magnitude distribution as the void galaxies. 
In Pan et al. (2011), we analysed the Sloan Digital Sky Survey Data
Release 7 and found a total of 1054 dynamically distinct voids
with radius larger than 10$h^{-1}$ Mpc. The voids are not empty, but are
underdense, with $\delta \rho / \rho < $-0.9 in their centers. 
In this paper we study the photometric properties of these void
galaxies. We look at the $u - r$ colours as an indication of star
formation activity and the inverse concentration index as an indication of
galaxy type. We find that void galaxies are statistically bluer than
galaxies found in higher density environments with the same magnitude
distribution. We examine the colours of the galaxies as a function of
magnitude, dividing the galaxies into bright, medium, faint and dwarf
groups, and we fit each colour distribution with a double-Gaussian
model for the red and blue subpopulations.
As we move from bright to dwarf galaxies, the population of red
galaxies steadily decreases and the fraction of blue galaxies
increases in both voids and walls, however the fraction of blue galaxies in the 
voids is always higher and bluer than in the walls.
We also split the void and wall galaxies into samples depending on
galaxy type, as measured by the inverse concentration index. We find that late type void galaxies are bluer
than late type wall galaxies and the same holds for early galaxies. 
We also find that early type, dwarf void galaxies are blue in colour.  
We also study the properties of void galaxies as a
function of their distance from the center of the void. We find very
little variation in the properties, such as magnitude, colour and type, 
of void galaxies as a function of their location in the void. The only
exception is that the dwarf void galaxies may live closer to the
centers of voids. As shown by Pan et al. (2011), the centers of voids
have very similar density contrast and hence all void galaxies live in
very similar density environments, which may explain the lack of
variation of galaxy properties with location within voids.

\end{abstract}


\section{Introduction}

The completion of the Sloan Digital Sky Survey allows identification of 
$\sim 10^3$ large voids and a sample of $\sim 10^5$ galaxies that lie 
within the centers of the voids.
With this sample we can study the photometric properties of
void galaxies within specific ranges of magnitude, colour,  surface
brightness profiles and distance from the center of the void. 
This void galaxy sample includes
$\sim 10^3$ galaxies with r-band magnitude $> -17$, thus the
properties of dwarf void galaxies can be studied. 

The prominence of voids in the large-scale distribution of galaxies
was brought to the attention of most astronomers by the discovery of a
void with diameter of $50h^{-1}$ Mpc in the direction of B\"{o}otes
\citep{Kirshner:1981}.
Subsequent surveys that were larger in both areal coverage and number density of galaxies
showed that voids are an important part of the cosmic web and fill more than half the volume of the universe
\citep{Geller:1989, Pellegrini:1989, daCosta:1988, Shectman:1996}.
Until recently, finding large voids has been challenging: voids have
diameters up to 60 h$^{-1}$ Mpc, which is similar to the minimum
dimension of early galaxy redshift surveys. The advent of 
the 2 degree Field Galaxy Redshift Survey and the Sloan
Digital Sky Survey has allowed samples of $\sim 10^3$ large voids to be identified
\citep{El-Ad:1997a, Muller:2000, Plionis:2002, Hoyle:2002, Hoyle:2004,
Ceccarelli:2006, Tikhonov:2006, Tikhonov:2007, Foster:2009, Pan:2011}.

The properties of objects within voids may provide strong tests of
models for cosmology and galaxy formation. \citet{Peebles:2001}
pointed out the apparent discrepancy between Cold Dark Matter Models
(CDM) and observations. CDM models predict a large population of low-mass halos
inside the voids \citep{Dekel:1986, Hoffman:1992}. 
However, pointed observations toward void
regions failed to detect a significant population of faint galaxies
\citep{Kuhn:1997,  Popescu:1997, McLin:2002}. Surveys of dwarf
galaxies indicate that they trace the same
overall structures as larger galaxies
\citep{Bingelli:1989}. \citet{Thuan:1987, Babul:1990} and \citep{Mo:1994} showed 
that galaxies have common voids regardless of Hubble type. 

\citet{Grogin:1999, Grogin:2000} identified a sample of 149 galaxies that
lie in voids traced by the Center for Astrophysics Survey.
Grogin \& Geller showed that the void galaxies tended to be bluer and
of later type. Their sample of 149
void galaxies covered a narrow range of absolute magnitudes ($-20 \le B \le -17$). 
Forty nine of their galaxies resided in a low density
contrast region with $\delta \rho /\rho \le 0.5$.

\citet{Rojas:2004} defined a sample of underdense galaxies from the
Sloan Digital Sky Survey Data Release 4. They used a nearest neighbour
technique to identify galaxies that had fewer than 3 neighbours in a
sphere of 7 h$^{-1}$Mpc. This corresponds to an underdensity of
$\delta \rho / \rho \le$ -0.6 around each void
galaxy, which is consistent with the local density in the interiors
of voids if void galaxies are clustered with $\xi(r)=1$ on this scale. 
Their sample contained $\sim 200$ galaxies in
their nearby sample (galaxies with $r$-band magnitudes between $-13.5$ and $-19.5$)
and $\sim 1000$ galaxies in their distant sample (galaxies with $r$-band
magnitude between $-17.5$ and $-21.5$) . 
They found that galaxies in underdense environments were bluer
than galaxies in higher density environments. They found that this
blueness was not explained by the morphology-density relation
\citep{Dressler:1980, Postman:1984}
as the nearby sample of both void and wall galaxies had similar
surface brightness profiles and yet the void galaxies were still
bluer. When the distant samples was split by galaxy type, the late
type and early type void galaxies were bluer than their counter-parts
at higher density. This result was confirmed by \citet{benda:2008} in the 2
degree Field Galaxy Redshift Survey.  

While we see some clear trends, controversy persists in the literature
as to whether or not galaxies in voids differ in their internal
properties from similar objects in denser regions. For example,
\citet{Rojas:2004, Blanton:2005, Patiri:2006, benda:2008} reach varying conclusions that
clearly depend on how environment is defined and which observed
properties are compared. There is also a marked difference between
properties of the least dense 30\% of galaxies (in regions with
density contrast $\delta < −0.5$ ) and objects in deep voids which form the
lowest density 10\% of galaxies (in regions of density contrast $\delta <
−0.9$, which is the theoretical prediction for the interiors of voids
that are now going non-linear). It is important to clearly define 
what “void” means; samples of “isolated” galaxies
selected by visual inspection (e.g., \citep{Karachentseva:2009} are
not necessarily in large-scale voids.

The aim of this paper is to analyze the properties of a large sample
of galaxies that actually reside in dynamically distinct voids (the
radial profiles of the voids match the predictions of void growth by
gravitational instability (see \citep{Pan:2011}) as opposed to smaller
scale underdense environments.  This sample of 88,794 galaxies is also
much larger than any other sample. For example there are as many voids
in the current sample as there were void galaxies in the sample of
\citet{Rojas:2004}. This allows us to compare the properties of void
galaxies with galaxies in higher density environments for well defined
ranges of magnitude, color, surface brightness profile and distance
from void center. We also have 4,500 galaxies with $M_r \> -17$ so we
can analyse the properties of dwarf galaxies in the voids as a
separate sample.

We organize the paper is as follows. In section ~\ref{sec:data} we define
data samples we will use. In ~\ref{sec:photoprops} we describe
the properties we consider and in section ~\ref{sec:results} and
~\ref{sec:concs} we describe the results and conclude our
findings.

\section{Data}
\label{sec:data}

\subsection{The Sloan Digital Sky Survey}

The SDSS is a wide-field photometric and spectroscopic survey
that covers 10$^4$ square degrees, including CCD
imaging of 10$^8$ galaxies in five colors and follow-up spectroscopy of
10$^6$ galaxies with $r < 17.77$. \citet{York:2000}
provides an overview of the SDSS and \citet{Stoughton:2002} describes
the early data release (EDR) and details about the photometric and
spectroscopic measurements made from the data. \citet{Abazajian:2003}
describes the First Data Release (DR1) of the SDSS. Technical articles
providing details of the SDSS include descriptions of the photometric
camera \citep{Gunn:1998}, photometric analysis \citep{Lupton:2001}, the
photometric system \citep{Fukugita:1996, Smith:2002}, the
photometric monitor \citep{Hogg:2001}, astrometric calibration
\citep{Pier:2003}, selection of the galaxy spectroscopic samples \citep{Strauss:2002,
  Eisenstein:2001}, and spectroscopic tiling \citep{Blanton:2001}. 

In this study we use the Korea Institute for Advanced Study Value-Added Galaxy
Catalog (KIAS-VAGC) \citep{Choi:2010} which is based on the SDSS Data
Release 7 (DR7) sample of galaxies \citep{Abazajian:2009}. The main source
of galaxies is the New York University Value-Added Galaxy Catalog (NYU-VAGC) Large Scale
Structure Sample (brvoid0) \citep{Blanton:2005} which includes
583,946 galaxies with $10 < m_r \le 17.6$ taken from DR7. 
929 objects are removed from the sample as they are mostly 
de-blended outlying parts of large galaxies. 
In the KIAS-VAGC, 10,497 bright galaxies
are added {\it into} the sample because they were too bright to be observed in 
the SDSS spectroscopic sample. These extra
galaxies have redshifts observed earlier by the UZC,
PSCz, RC3, or 2dF surveys. In the KIAS-VAGC, an additional 114,303 galaxies with $17.6 < m_r
< 17.77$ are included from the NYU-VAGC (full0). This yields a total of 707,817
galaxies in the parent sample that we examine. This catalog offers an extended magnitude range with high
completeness over apparent magnitude range $10 < m_r < 17.6$. 

\subsection{Construction of the Void Galaxy Sample}

We use a sample of voids found within a volume limited sample
constructed from the parent catalog described above. The redshift
limit we chose for finding voids was $z_{max}=0.107$, which corresponds to an absolute magnitude limit of
$M_r<-20.09$. This sample contains 120,606 galaxies, which is the
maximum number of galaxies possible in a volume limited sample. See
\citet{Pan:2011} for full details of this choice of volume limit.
The resulting sample of voids found by \citet{Pan:2011} includes
a total of 1054 voids with radius as large as 25$h^{-1}$Mpc. For
full details see \citet{Pan:2011} and \citet{Hoyle:2002}; here we give a
brief description of the most important aspects of VoidFinder.

The VoidFinder algorithm uses the coordinates of
the galaxies in a volume limited sample to find statistically
significant cosmic voids. It effectively finds large voids of density
contrast $\delta \rho / \rho < 0.9$ in the center and
radius R $>10h^{−1}$Mpc. VoidFinder is
based on the \citet{El-Ad:1997b} method for finding voids and
was used to find voids in galaxy redshift surveys by
\citet{Hoyle:2002}, \citet{Hoyle:2004}, and \citet{Pan:2011}. VoidFinder has been
applied to many surveys, including the
IRAS PSCz, CfA2+SSRS2, 2dF, SDSS, and 6dF. In cases where there is overlap
between the different surveys, we have consistently found similar voids. Our tests on
cosmological simulations demonstrated that this method
works in identifying voids in the distributions of both simulated
galaxies and dark matter \citep{Benson:2003a}.

The first step of the algoritm is to identify galaxies in the
volume-limited sample that lie in low-density environments using a
nearest neighbour analysis in 3D. If a galaxy has less than three
neighbours in a sphere of $6.3h^{-1}$ Mpc it is classified as a field
galaxy. This threshold is chosen because in this sample it corresponds
to local galaxy density $\delta \rho / \rho < -0.6$, which is
consistent with voids of density 
($\delta \rho / \rho < -0.8$) if void galaxies are clustered with
$\xi(r)=1$ on this scale (i.e., clustering of galaxies raises the
density measured around a galaxy above the mean). These field galaxies
are temporarily removed from the galaxy sample. The rest of the
galaxies are then placed on a grid and void finder searches for empty
grid cells. The grid cell size we use is 5 $h^{-1}$ in length which
guarantees that all spheres larger than 8.5 $h^{-1}$ Mpc in radius are
detected by the algorithm. A maximal sphere is grown from each empty
cell, but the center of the maximal sphere is not confined to the
initial cell. Eventually the sphere will be bound by 4 wall galaxies.
There is redundancy in the detecting of maximal spheres, but this is
useful to define non-spherical voids. The largest empty sphere is the
basis of the first void region.  If there is an overlap of $> 10\%$
between an empty sphere and an already defined void then the empty
sphere is considered to be a subregion of the void, otherwise the
sphere becomes the basis of a new void.  There is a cutoff of 10
$h^{-1}$ Mpc for the minimum radius of a void region as we seek to
find large scale structure voids that are dynamically distinct and not
small pockets of empty space created by a sparse sample of
galaxies. For further details of this implementation of the VoidFinder algorithm see
\citet{Hoyle:2002}.

\begin{figure}
  \centering
  \includegraphics[width=2.0in,angle=270]{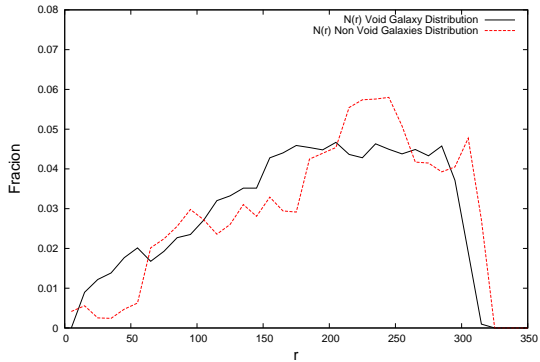}
  \caption{Radial distribution of all SDSS galaxies with z$<$0.107 (red,
    dashed line) and void galaxies  
    (black, solid line). The two distributions follow a similar trend
    but are not exactly the same because voids are constrained to lie
    completely within the survey boundaries.}
  \label{rdist}
\end{figure}

The final steps are to identify the void galaxies and form a luminosity-matched wall galaxy sample. 
We use the full apparent magnitude limited sample and simply check if the coordinates of
every galaxy lie within a void or not. This creates the void
sample. 
A void galaxy does not have to lie
in the central maximal sphere; it may lie in one of the outer spheres
that make up the total void volume. 
We find a total of 88,794 void galaxies that lie in
void regions with density contrast $\delta \rho /
\rho <$ -0.9, which yields the largest and most underdense sample of void
galaxies to date. The radial and magnitude distributions of the void
galaxies are shown in figure ~\ref{rdist} and ~\ref{fig:wallnmagdist} 
(black line in both cases). Note that
void galaxies are not found at the minimum or maximum distance of the
volume limited sample because the void galaxies are required to live within
a void and the voids are required to be within the survey, thus the void
galaxies have a flatter radial distribution, $n(r)$, than all galaxies in the
magnitude limited sample. Figure ~\ref{fig:wallnmagdist} shows the wide
range of absolute magnitudes of void galaxies, $-23 < M_r < -12$.

In our previous analysis, \citep{Rojas:2004} the SDSS was incomplete and
there were only $\sim 200$ galaxies in the nearby faint sample (with
r-band magnitudes in the range $-13.5$ to $-19.5$), and $\sim 1000$ galaxies in the
distant sample (r-band magnitudes $-17.5$ to $-21.5$). 
Also, the manner in which galaxies were classified as void
galaxies was different. In that earlier study, due to the limited volume of the
SDSS at that point, galaxies were classified as void galaxies solely using a
nearest neighbour analysis.
Now that the SDSS in complete we have 88,794 void
galaxies, all of which lie in bona fide voids.
The tremendous increase in void galaxy sample size allows us
to study the properties of void galaxies with greater precision and/or
allows us to carefully define the range of parameters for comparison
between void and wall galaxies. 

\subsection{Construction of the Wall Galaxy sample}
\label{sec:wallgalsample}

\begin{figure}
\centering
 \includegraphics[width=2.0in,angle=270]{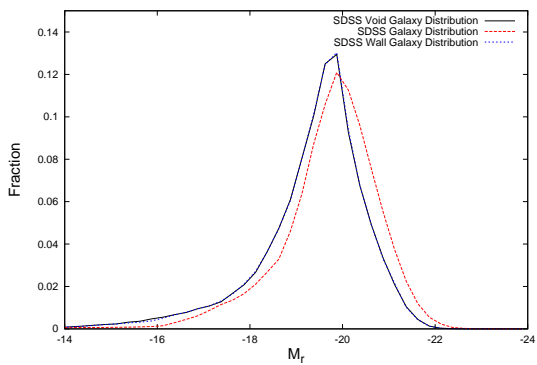}
  \caption{Fraction of galaxies as a function of their r-band absolute
    magnitude for all galaxies (red, dashed), void galaxies (black, solid) and the
    sparse-sampled set of wall galaxies (blue, dotted). By construction, this
    sample of wall galaxies has the same $r$-band absolute magnitude
    distribution as the void galaxies. }
  \label{fig:wallnmagdist}
\end{figure}

Figure ~\ref{fig:wallnmagdist} shows that the distribution of absolute
magnitudes presents
a clear shift toward fainter magnitudes in voids, consistent
with our previous work \citep{Hoyle:2005}. In \citet{Goldberg:2005}
we showed that the mass function shifts to lower masses in voids. In
this paper we want to compare the properties of void galaxies and wall
galaxies with the sample absolute magnitude.  

In all comparisons between void and wall galaxies that are presented
below, we examine a sample of wall (non-void) galaxies that have been
sparse sampled in such a way as to match the absolute magnitude
distribution of the void galaxies The r-band absolute magnitude
distribution of the sparse-sampled wall galaxy sample is shown in
figure ~\ref{fig:wallnmagdist}. It can be seen that the void galaxies
(black, solid) and the wall galaxies (blue, dotted) now have the same
distributions, thus comparisons between them will reveal differences
that cannot be attributed to differences in luminosity alone.

\subsection{Samples}
In order to check for dependence on luminosity of any of the results,
we split both the wall galaxies and void galaxies into samples of
absolute magnitude. These are defined in table ~\ref{table:sample}. ``Bright''
galaxies have M$_r <$-19.9, the ``Medium'' samples have -19.9$<$M$_r
\le$-19.2, the ``Faint'' samples have -19.2$<$M$_r \le$-17.0 and the
``Dwarf'' sample has M$_r > $ -17.0. The limits between bright, medium
and faint are chosen so that approximately 28,000 galaxies are in
each sample. The dwarf sample is defined as galaxies with magnitude 
less than -17 so that we are looking at only the faintest galaxies. There are roughly
4,500 galaxies in this sample. The exact number of void and wall
galaxies in each sample is given in table ~\ref{table:sample}.

\begin{table} 
\centering 
\begin{tabular}{c c c}  
Name & Magnitude Range & N(Galaxies) \\
\hline 
Void & All & 88794 \\
Void Bright & M$_r <$-19.9 & 29275 \\
Void Medium & -19.9$<$M$_r \le$-19.2 & 28743 \\
Void Faint & -19.2$<$M$_r \le$-17.0 & 26199 \\
Void Dwarf & M$_r > $ -17.0 & 4577 \\
\hline
Wall & All & 88356 \\
Wall Bright & M$_r <$-19.9 & 29440 \\
Wall Medium & -19.9$<$M$_r \le$-19.2 &28560 \\
Wall Faint & -19.2$<$M$_r \le$-17.0 & 26005 \\
Wall Dwarf & M$_r > $ -17.0 & 4351 \\
\hline 
\end{tabular} 
\caption{The name, magnitude range and number of galaxies in each sample} 
\label{table:sample} 
\end{table}

\begin{figure}
\centering
\includegraphics[width=2.0in,angle=270]{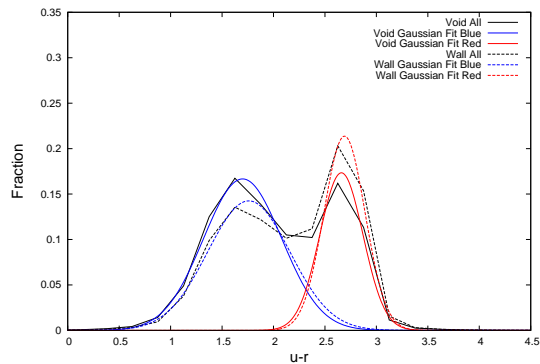}
\caption{The fraction of galaxies as a function of their u-r colour
for void galaxies (black, solid) and wall galaxies (black, dashed). This diagram
displays the characteristic split between blue, spiral type galaxies
and red, elliptical galaxies. There are more elliptical galaxies in the walls and more
spirals in the voids. To demonstrate this more clearly, we show the
double-Gaussian fits to the red and blue portions of the $u - r$ distribution. The
red dashed line (wall) is higher than the red solid line (void),
while the blue solid line (void) is higher than the blue dashed line (wall). }
\label{fig:ur}
\end{figure}

\section{Photometric Properties}
\label{sec:photoprops}

It is well known that red galaxies tend to populate regions of higher
density, such as clusters, and tend to be elliptical. Blue galaxies tend to be found in lower
density environments, tend to be spiral and are also known to be less
clustered than red galaxies \citep{Postman:1984, Dressler:1980,
Strateva:2001, Baldry:2003, Hogg:2002, Blanton:2002} 
This behavior is shown in the SDSS galaxy photometry by
\citet{Blanton:2002}; see their Figures 7 and 8, in 
which they find that the
distribution of $g-r$ colors at redshift $z=0.1$ is bimodal.

To compare the colours of void and wall galaxies, we consider the
$u-r$ colour, which is sensitive to the UV flux and the 4000 \AA
break. To compare morphological properties of void and wall galaxies,
we examine the distribution of inverse concentration indices measured
by the SDSS photometric pipeline \citep{Lupton:2001, Stoughton:2002,
Pier:2003}. The inverse concentration index (ICI) is defined by the
ratio ICI $=r_{50}/r_{90}$, where $r_{50}$ and $r_{90}$ correspond to the radii at
which the integrated fluxes are equal to 50\% and 90\% of the
Petrosian flux, respectively. A small value of ICI corresponds to a
relatively diffuse galaxy and a large value of ICI to a highly
concentrated galaxy. The concentration index (or the inverse
concentration index) has been shown to correlate well with galaxy type
\citep{Strateva:2001, Shimasaku:2001}.

\subsection{Distance from Void Center}
\label{sec:distancevoidcent}

An additional parameter we consider is the distance of the void galaxy from
the center of the void. To find the true centers of void regions, we first grid the
space into cubes $0.5 h^{-1}$Mpc on a side.  We identify all cubes whose centers lie
within the void region using the set of holes that compose the void
region.  Each cube is given equal weighting, and the center of mass
of the set of void cubes is found.  This center of mass is used as
the center of the void region rather than the center of the maximal
sphere but these two positions are very close together (Pan 2012, in prep). 
The distance of each void galaxy from this position is then
calculated. We also calculate the distance as a
fraction of the effective radius of each void.

\begin{figure*}
\centering
\begin{tabular}{cc}
\includegraphics[width=2.5in,angle=270]{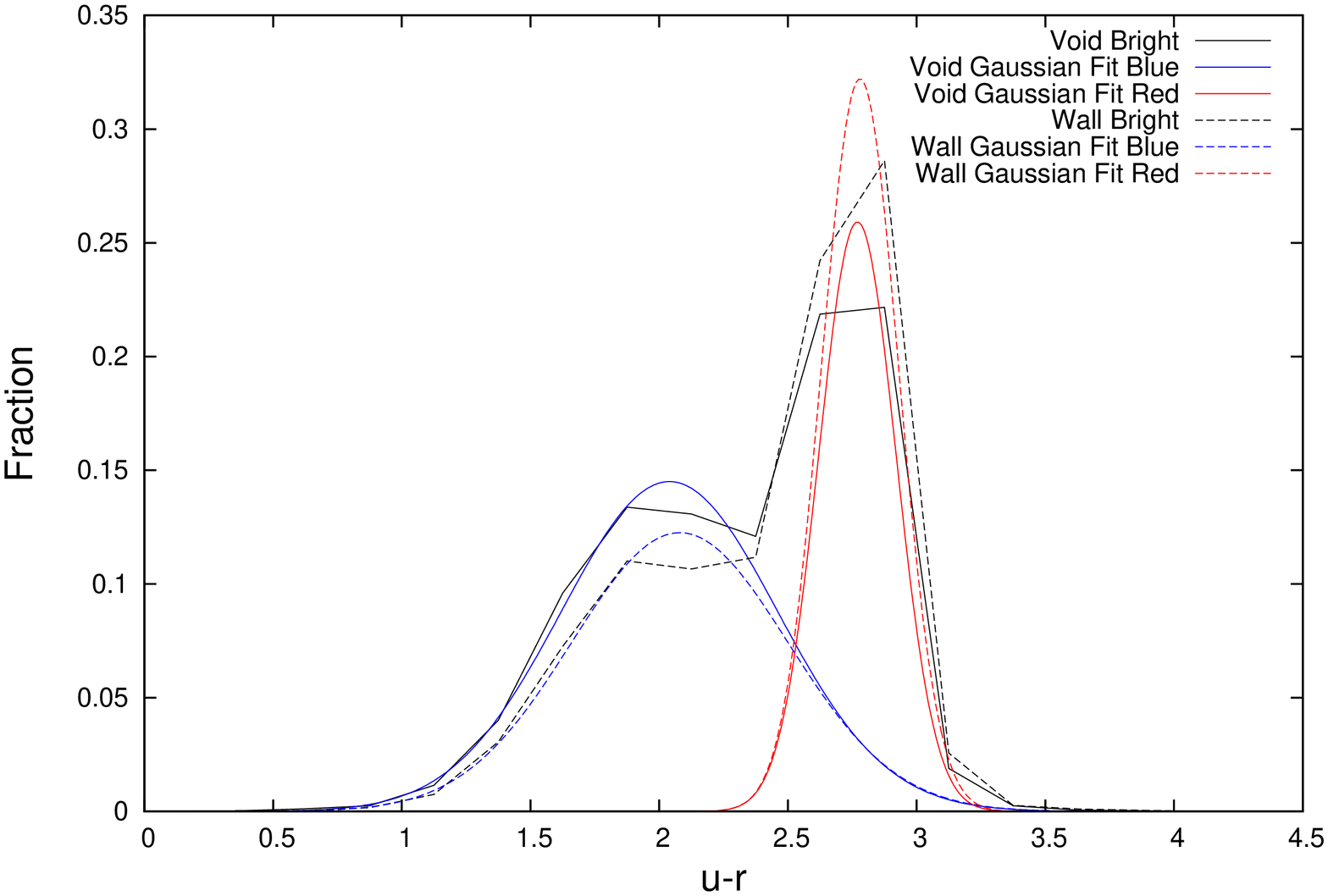}
& 
\includegraphics[width=2.5in,angle=270]{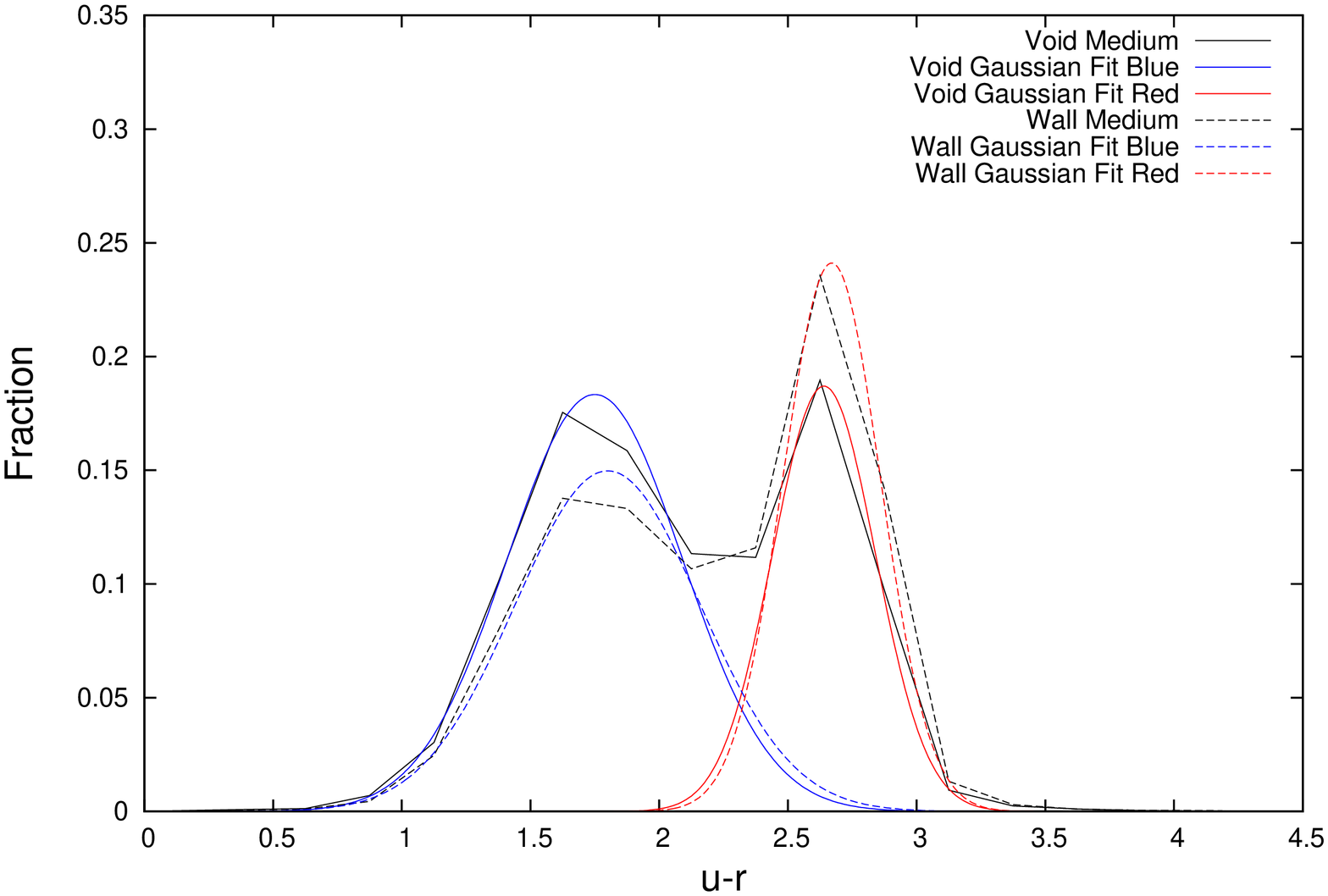}
\\
\includegraphics[width=2.5in,angle=270]{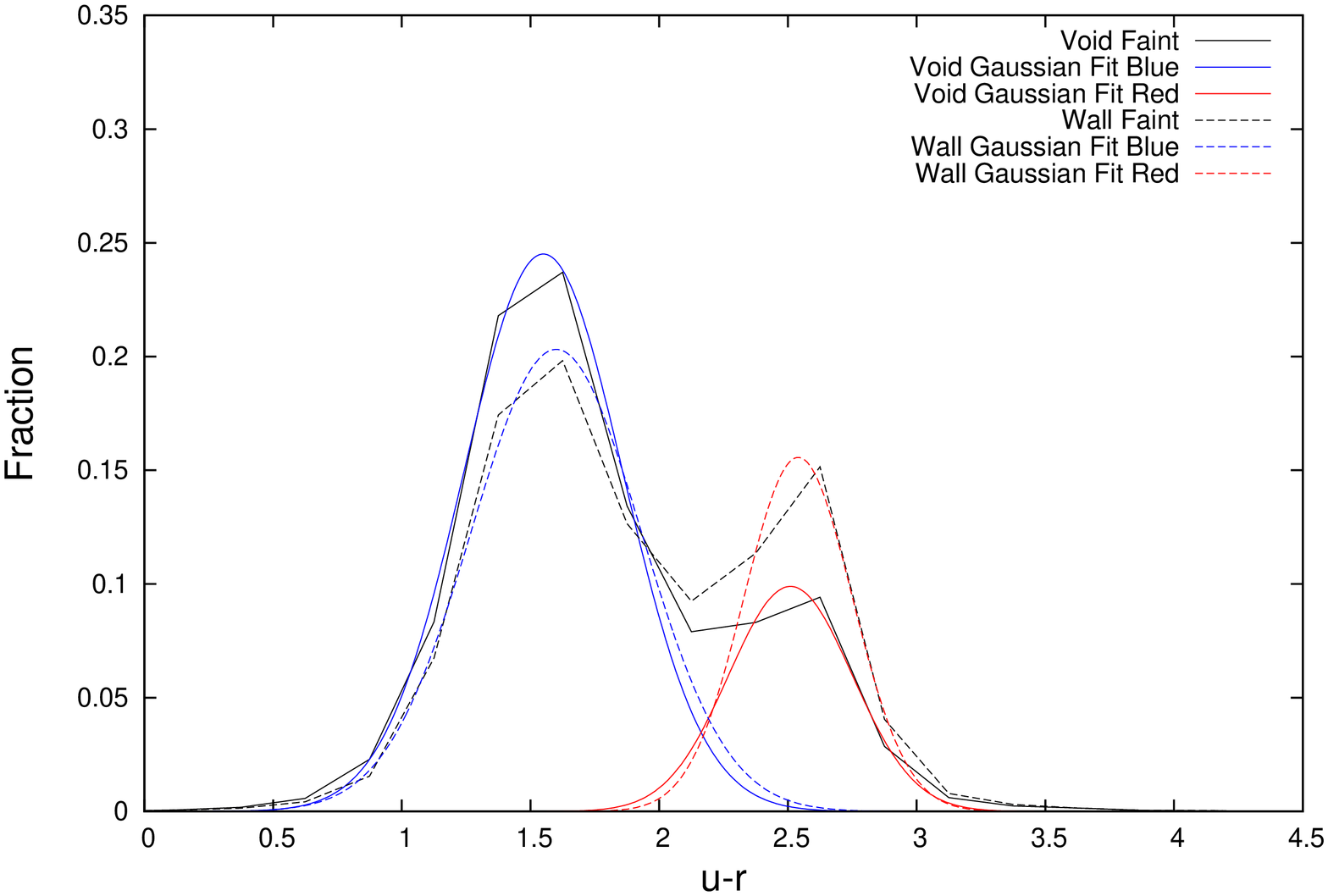}
&
\includegraphics[width=2.5in,angle=270]{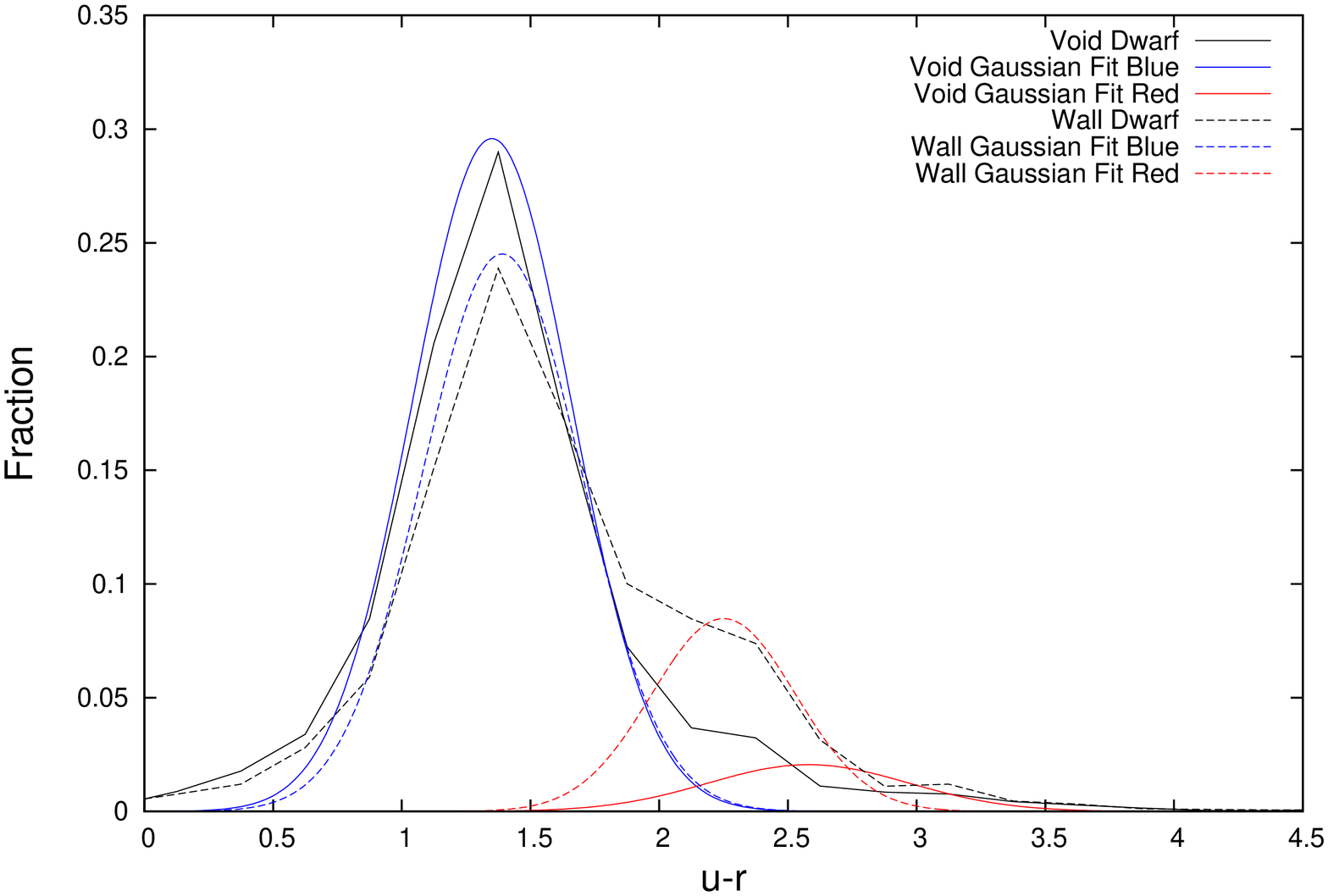} \\
\end{tabular}
\caption{The fraction of galaxies as a function of their $u - r$ colour
for void galaxies (black, solid) and wall galaxies (black, dashed) for
the bright (top, left), medium (top, right), faint (bottom, left) and
dwarf (bottom, right) samples of galaxies. As we move
from bright magnitudes to fainter magnitudes the mix of galaxies
changes. The proportion of red galaxies in the bright sample is
higher than in the dwarf sample and vice versa for blue galaxies. In
all plots, there are more red galaxies in the walls and more
blue galaxies in the voids. In all cases, the
red dashed line (wall) is higher than the red solid line (void), the
blue solid line (void) is higher than the blue dashed line (wall). }
\label{fig:urbrightdwarf}
\end{figure*}

\section{Results}
\label{sec:results}

\subsection{Colour Comparison}

In Figures ~\ref{fig:ur} and ~\ref{fig:urbrightdwarf} we show a 
comparison of the $u - r$ colours of void galaxies (black, solid) 
and wall galaxies (black, dashed). The plots show the
number of galaxies with particular $u - r$ colour divided by the total
number of galaxies in the sample. In all cases there is
a shift to the left for the void galaxies. This indicates that at all
magnitudes, void galaxies are bluer than wall galaxies. 

Figure ~\ref{fig:ur} can be compared to Figure 7, top left, of Rojas el
at. (2004). In both cases the void sample contains more blue galaxies and
the wall sample more red galaxies. The differences between the void
and wall samples are more significant in this work as both void and wall samples
now have the same absolute magnitude distribution and the void sample
contains 80 times more galaxies.

In table ~\ref{table:photomean}, for each sample we list the mean $u - r$
colour and the sample deviation, which is the standard deviation divided by the square root of the
total number of galaxies. The sample deviations are small due the large number of galaxies that we have available. 
As can be seen, the mean colour of bright void galaxies is more than
5$\sigma$ bluer than the brightest wall galaxies. 
As we move from brighter to fainter galaxies, both the void and wall
galaxy samples become bluer. The dwarf
void galaxies are also bluer than the dwarf wall galaxies, although due to
the smaller sample size the difference is not as significant. 

Looking at figure \ref{fig:urbrightdwarf}, it is clear that, in
addition to an overall shift in
colour between void and wall regions, there is also a change in
population from red galaxies to blue galaxies as the magnitude of the
sample becomes fainter. 
In the brightest sample there are more red galaxies
than blue in both void and wall samples. As the luminosity decreases,
the height of the red peak decreases and the blue peak grows. By the
time we reach the dwarf population, there are hardly any red
galaxies in the void sample and only a few red galaxies in the wall
sample. 

To better understand how the proportion of red and blue galaxies changes from sample
to sample, we fit a double Gaussian model to the $u - r$ distribution, as
suggested by  ~\citet{Baldry:2003}. One of the gaussians fits the red wing of the
colour distribution, the other the blue wing. There are six parameters
in the model, $\mu_{blue}$, $\mu_{red}$, $\sigma_{blue}$,
$\sigma_{red}$ and 
two normalisation factors, norm$_{blue}$ and norm$_{red}$.

\begin{equation}
g(x)_{blue} = \frac{1} {\sqrt{2 \pi \sigma_{blue}} }   e^{\frac{-(x -
    \mu_{blue})^2 }{2\sigma_{blue}^2} }
\end{equation}   
\label{gaussblue}
       
\begin{equation}
g(x)_{red} = \frac{1} {\sqrt{2 \pi \sigma_{red}} }   e^{\frac{-(x -
    \mu_{red})^2 }{2\sigma_{red}^2} }
\end{equation}   
\label{gaussred}             

Because the sum of the two Gaussian distributions is fixed by the total
number of galaxies in each sample, there are only 5 free parameters. 
We choose the normalisation of the red wing to be the fixed parameter.

\begin{equation}
\Sigma \left(\frac{g(x)_{blue}}{norm_{blue}} + \frac{g(x)_{red}}{norm_{red}}\right)  = 1
\end{equation}

To find the parameters, we use a coarse grid of the 5 free parameters
to get an estimate of the values. We then refine the search around the
best fit parameters. The values of $\mu$, $\sigma$ and the blue
normalisation are refined to $\pm 0.01$. 

We define the best fit as the one with the lowest value of 
\begin{equation}
\chi^2 = \Sigma \frac{(N_{data}(i) - (g_{blue}(i) +
  g_{red}(i))*N_{tot})^2} { ( (g_{blue}(i) + g_{red}(i))*N_{tot})}
\end{equation}
where $N_{tot}$ is the total number of galaxies in the sample. The
bins of $u - r$ are 0.25 in size. 

Because bins with very few galaxies can cause the value of $\chi^2$ to
be excessively large, we do not include any bin that contains less than 100 galaxies. We
do not apply any other smoothing to the distribution of galaxies. 

The double Gaussian fits to the $u - r$ distribution are 
shown in Figures ~\ref{fig:ur} and ~\ref{fig:urbrightdwarf} and the
values are listed in Table ~\ref{tab:gaussian}. As the 
galaxies change from bright to dwarf, the fraction
of blue galaxies increases in both the wall and void samples. 
These fits also show that the proportion of blue galaxies is higher in the
voids than in the walls in the full sample and in each of the samples split by
magnitude (the red dashed line is higher than the red solid line and
the blue solid line is higher than the blue dashed line). The void
galaxies are still bluer than the wall galaxies as the values of the
mean blue and red gaussians are bluer in the voids than in the walls 
in each of the gaussian fits, apart from for the red wing of the dwarf
sample, where there are very few galaxies to confine the fit.

\subsection{Inverse Concentration Index Comparison}

\begin{figure}
\centering
\includegraphics[width=2.0in,angle=270]{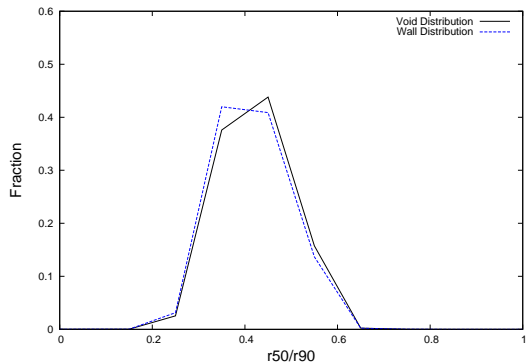}
\caption{The fraction of galaxies as a function of their inverse
 concentration index. Elliptical galaxies have low values of the
 inverse concentration index, spiral galaxies have higher values of
 the inverse concentration index. There is little visual difference 
between the curves for the void (black, solid) and wall
(blue, dashed) samples, although the wall galaxies statistically lower values than the
void galaxies, as shown in table ~\ref{table:photomean}.}
\label{conx}
\end{figure}

\begin{figure}
\centering
\begin{tabular}{cc}
\includegraphics[width=1.0in,angle=270]{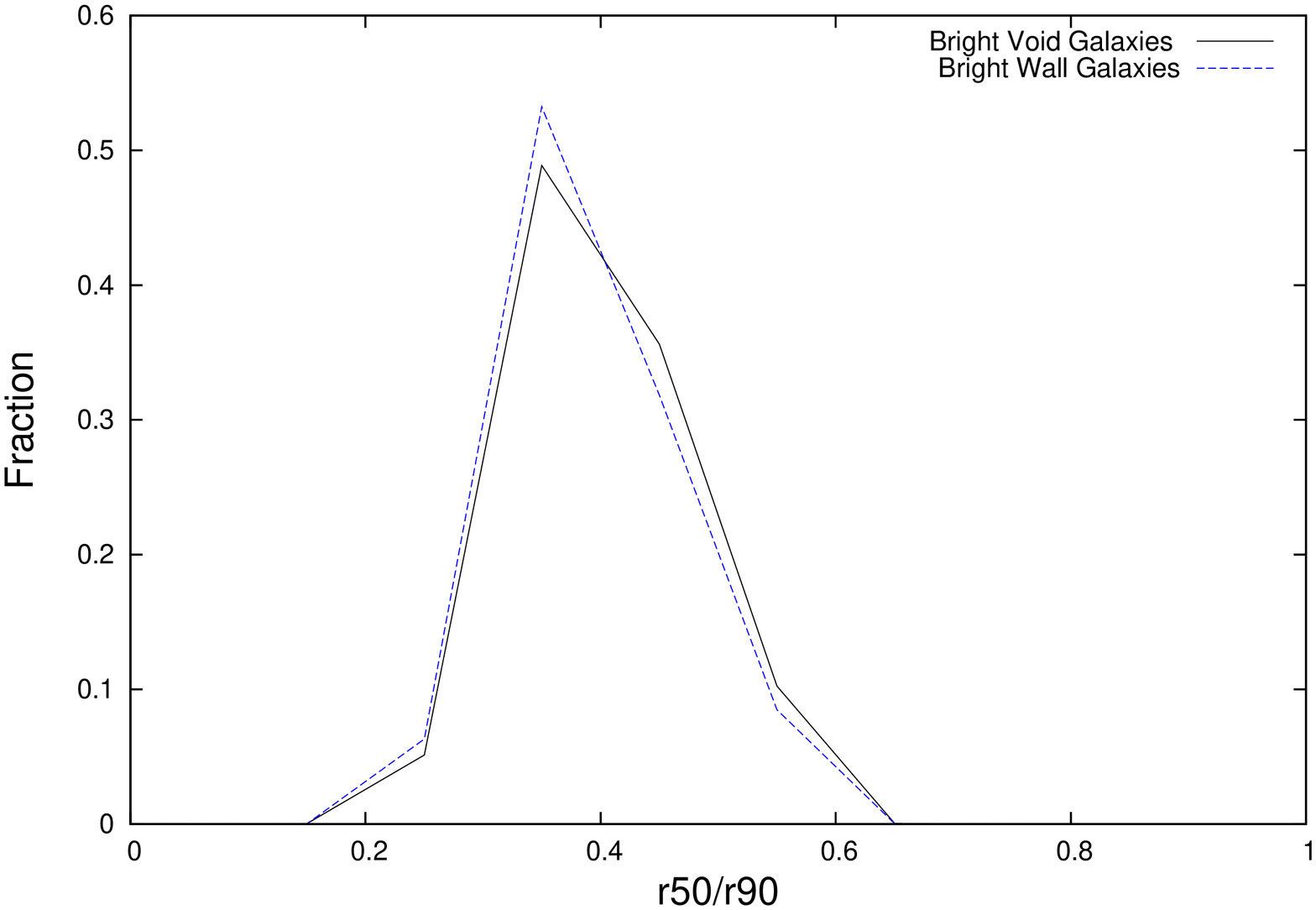}
& 
\includegraphics[width=1.0in,angle=270]{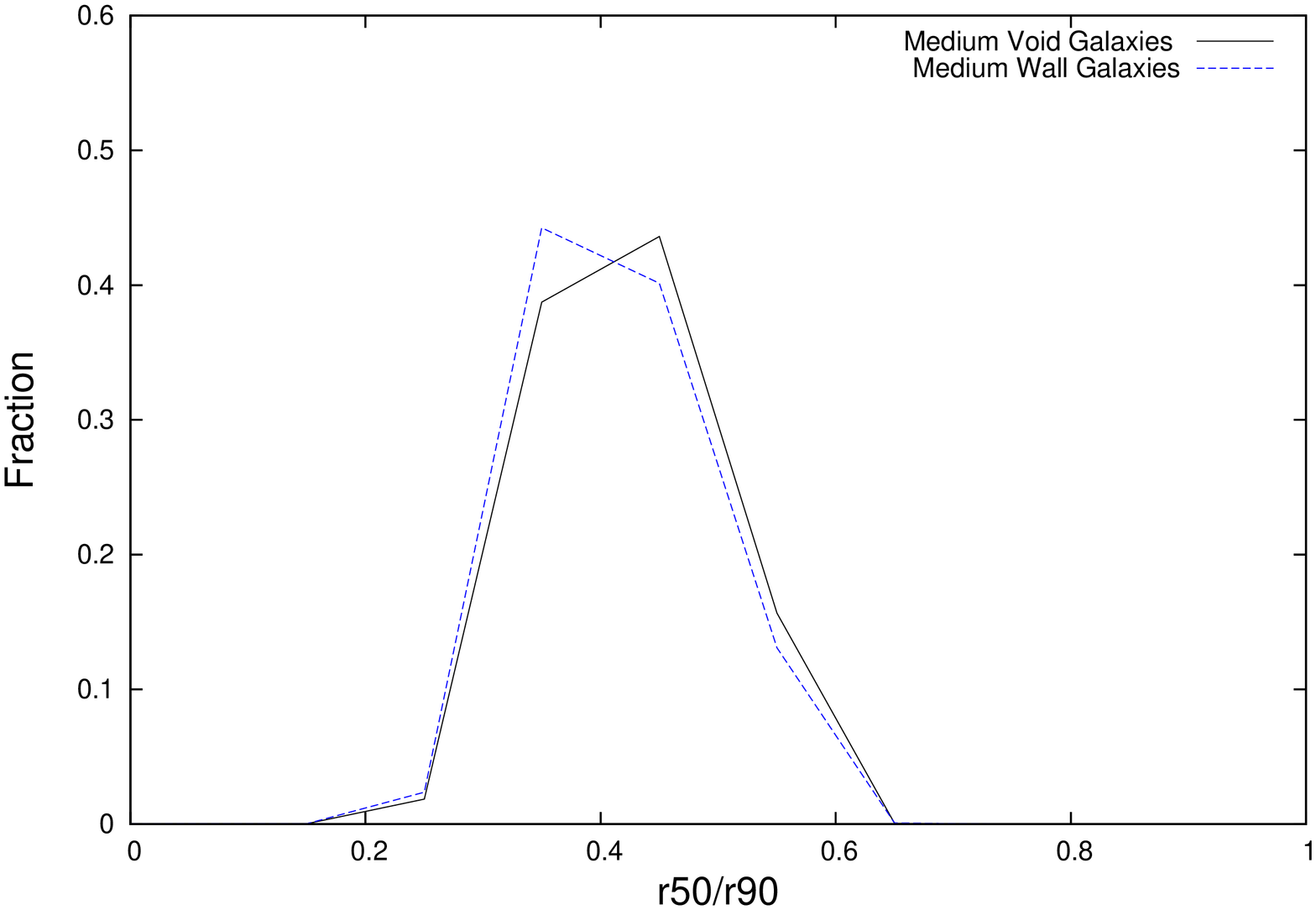}
\\
 \includegraphics[width=1.0in,angle=270]{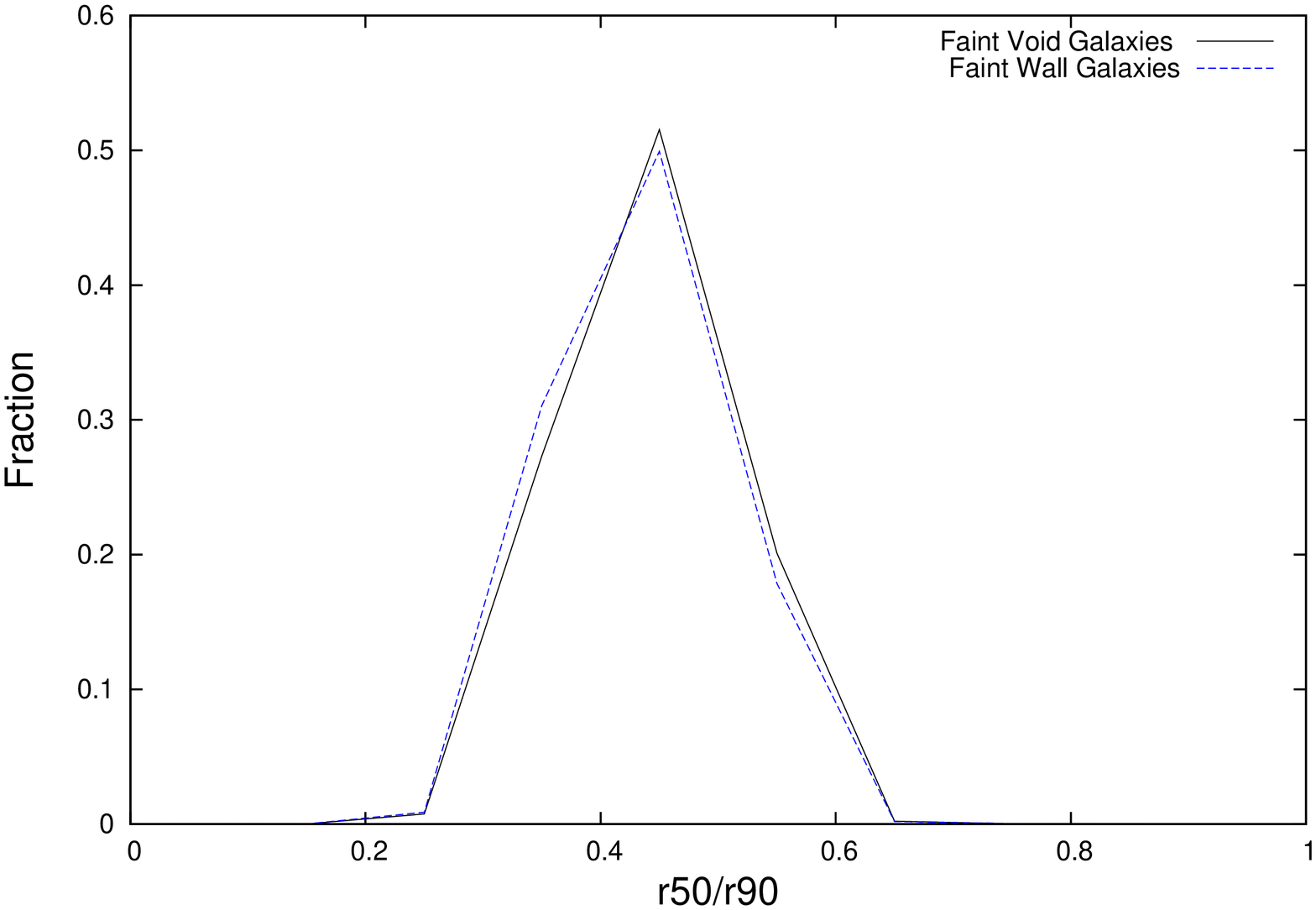}
 &
 \includegraphics[width=1.0in,angle=270]{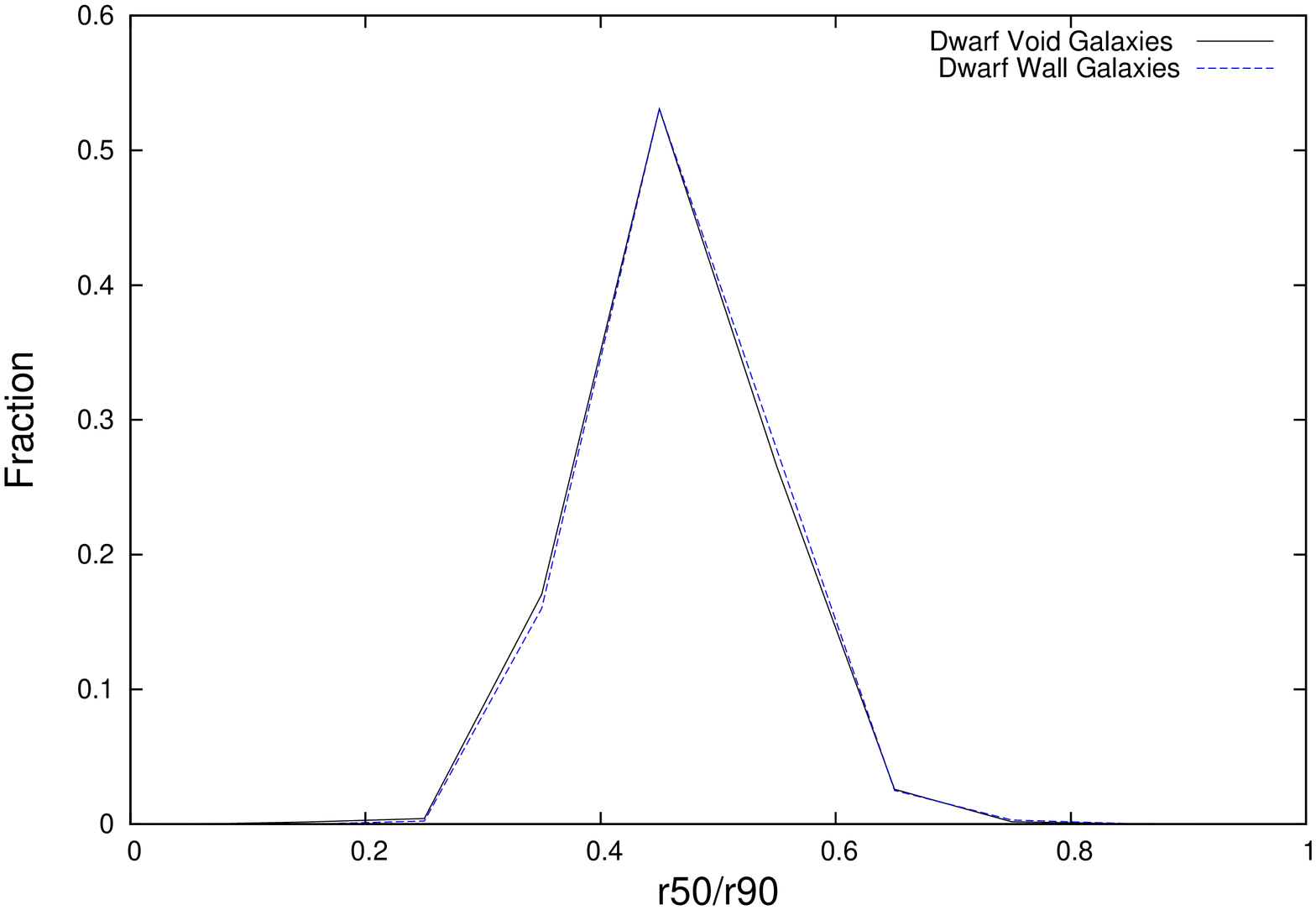} \\
\end{tabular}
\caption{The fraction of galaxies as a function of their inverse
  concentration index
 for void galaxies (black, solid) and wall galaxies (blue, dashed) for
 the bright, medium, faint and dwarf samples as indicated in the
 figure. There is little visual difference 
between the values for the void and wall samples but the wall galaxies
do have lower values of the ICI than the
void galaxies as shown in table ~\ref{table:photomean}. 
As we moved to fainter magnitudes, the galaxies have
higher inverse concentration indices, representing more late type galaxies. }
\label{fig:conxbrightdwarf}
\end{figure}

In Figure ~\ref{conx} we show the normalised histogram of inverse
concentration indices for the void galaxies (black, solid) and wall
galaxies (blue, dashed) for all the void and wall galaxies.  Figure
~\ref{fig:conxbrightdwarf} shows results for the samples as a function
of magnitude. It is clear that there is very little difference in the
types of galaxies within each magnitude bin; the black solid and
blue dashed lines are very similar. What can be seen is that as the
magnitude of the galaxies shifts from bright to faint, the mix of
galaxies changes. In the bright sample there are more early type
galaxies and in the dwarf samples there are more late type
galaxies. This is reflected in table ~\ref{table:photomean} as the
mean value of the inverse concentration index shifts from 0.3980 to
0.464 in the voids and from 0.3889 to 0.466 in the walls. This
means that when we are comparing void and wall galaxies in the bright,
medium, faint and dwarf samples, we are comparing galaxies with the
{\it same magnitude and galaxy type.}

\subsection{Colour as a function of Type}

\begin{figure}
\centering
\begin{tabular}{cc}
\includegraphics[width=1.0in,angle=270]{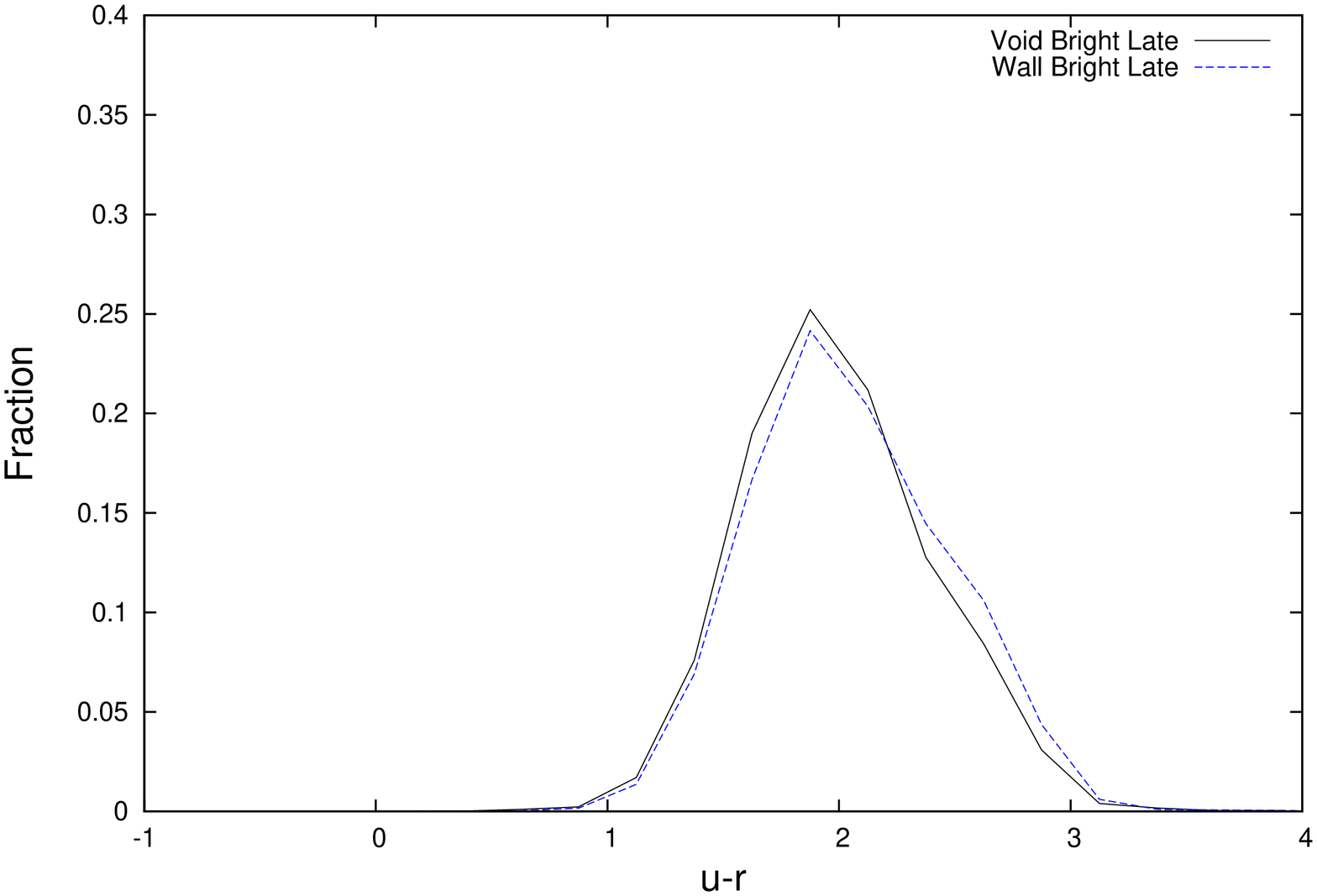}
& 
\includegraphics[width=1.0in,angle=270]{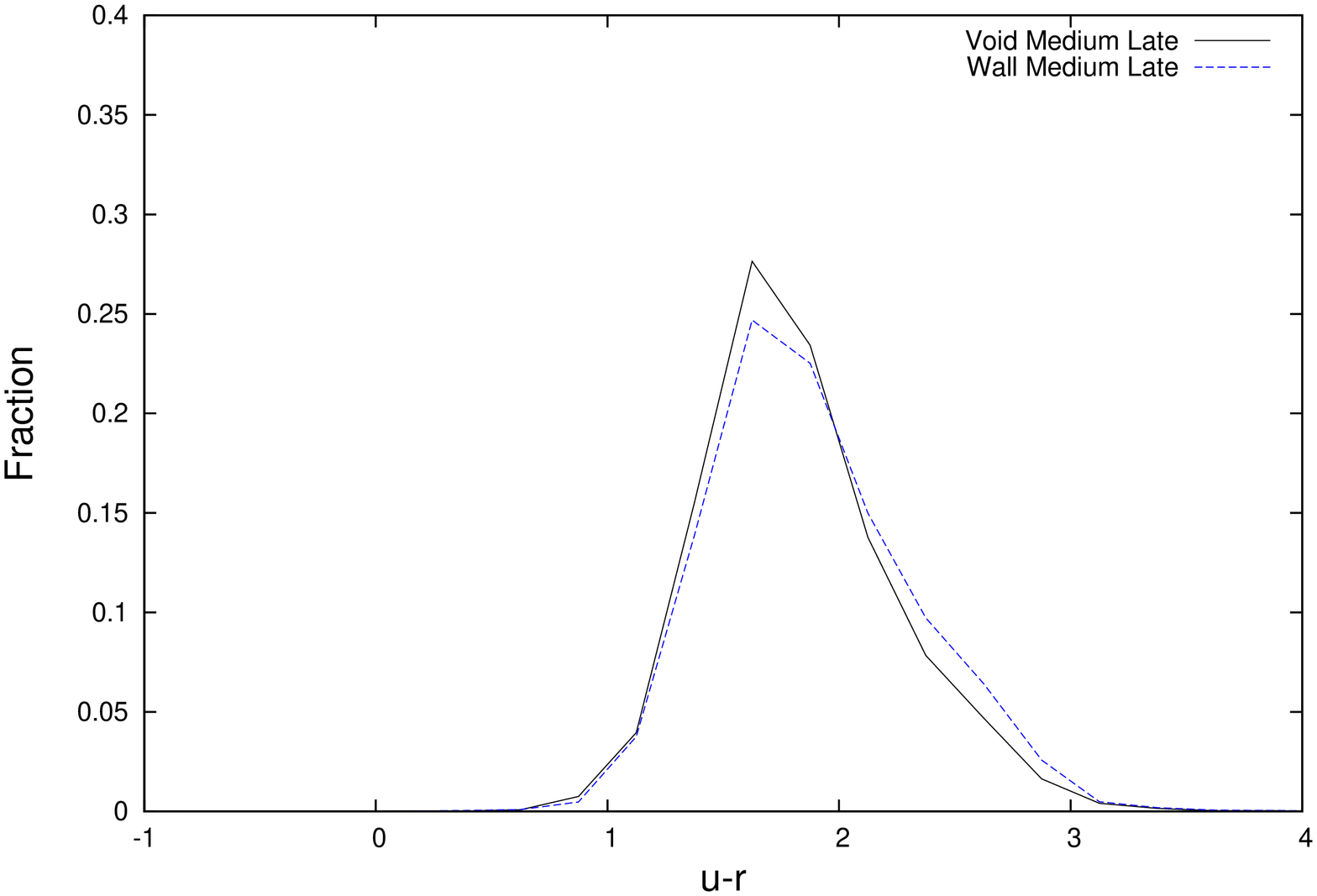}
\\
 \includegraphics[width=1.0in,angle=270]{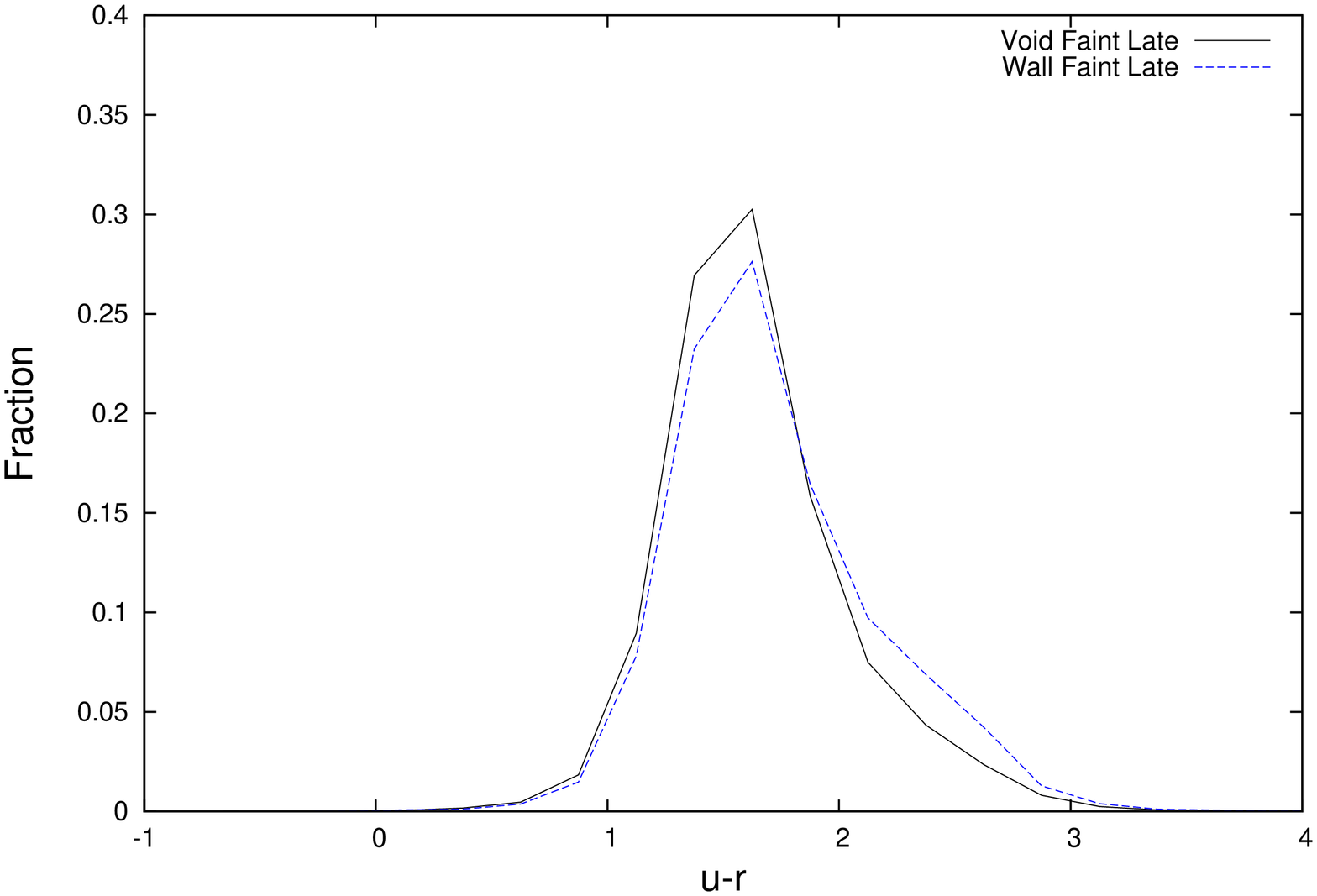}
 &
 \includegraphics[width=1.0in,angle=270]{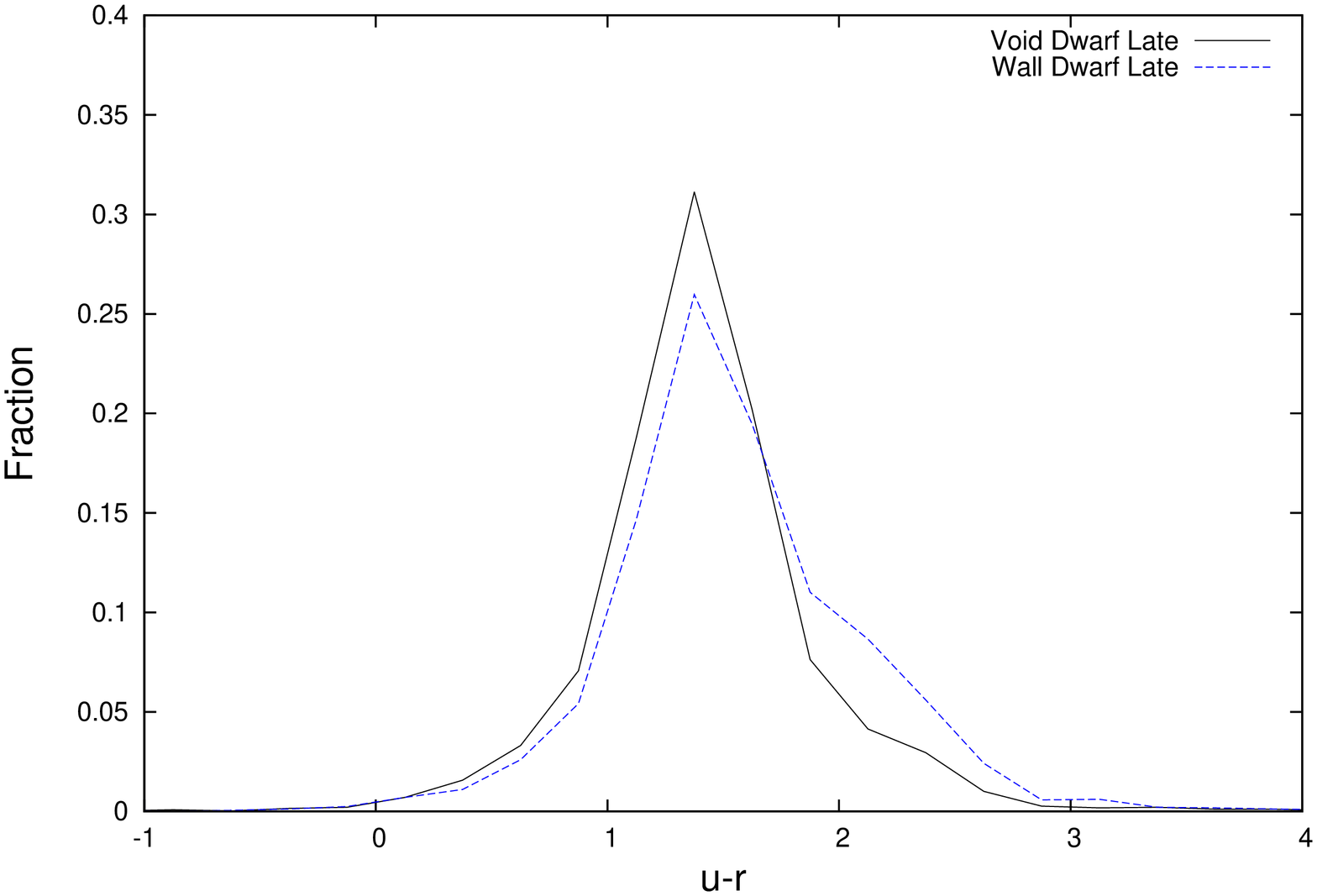} \\
\end{tabular}
\caption{The fraction of galaxies as a function of their $u - r$ colour 
 for late type (ICI $>$ 0.42) void galaxies (black, solid) and wall galaxies
 (blue, dashed) as a function of magnitude, as indicated in each figure. }
\label{fig:urspiral}
\end{figure}

\begin{figure}
\centering
\begin{tabular}{cc}
\includegraphics[width=1.0in,angle=270]{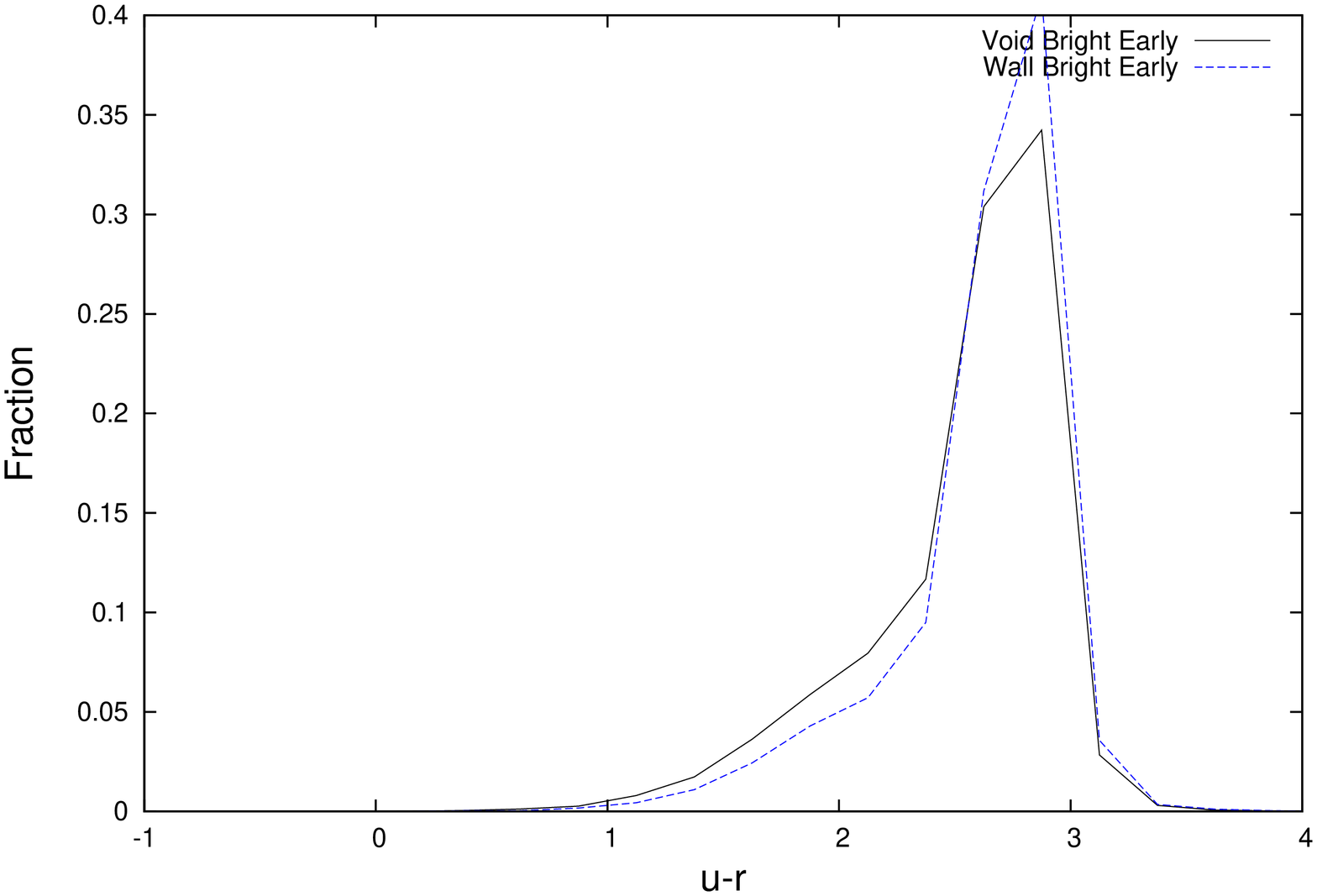}
& 
\includegraphics[width=1.0in,angle=270]{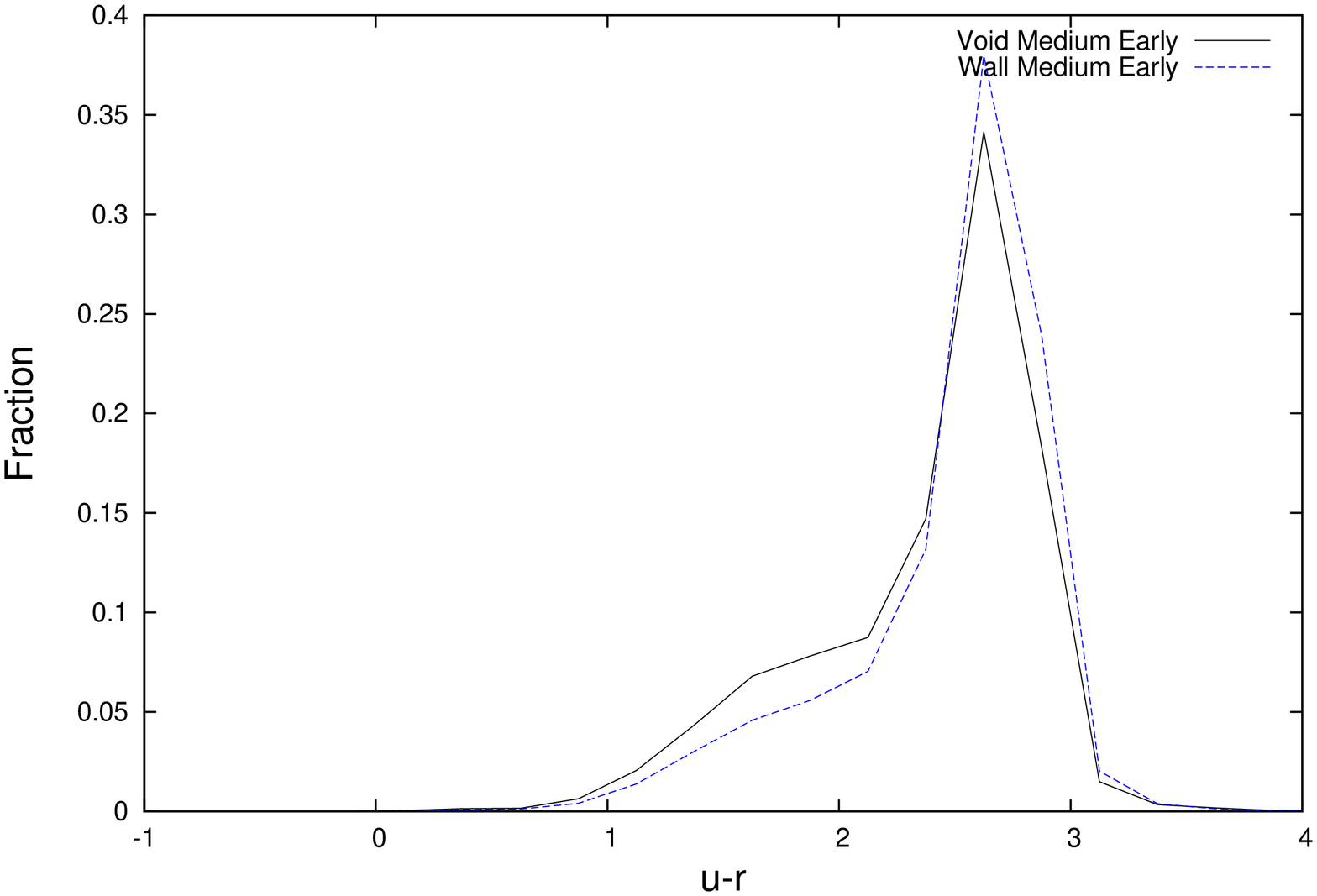}
\\
\includegraphics[width=1.0in,angle=270]{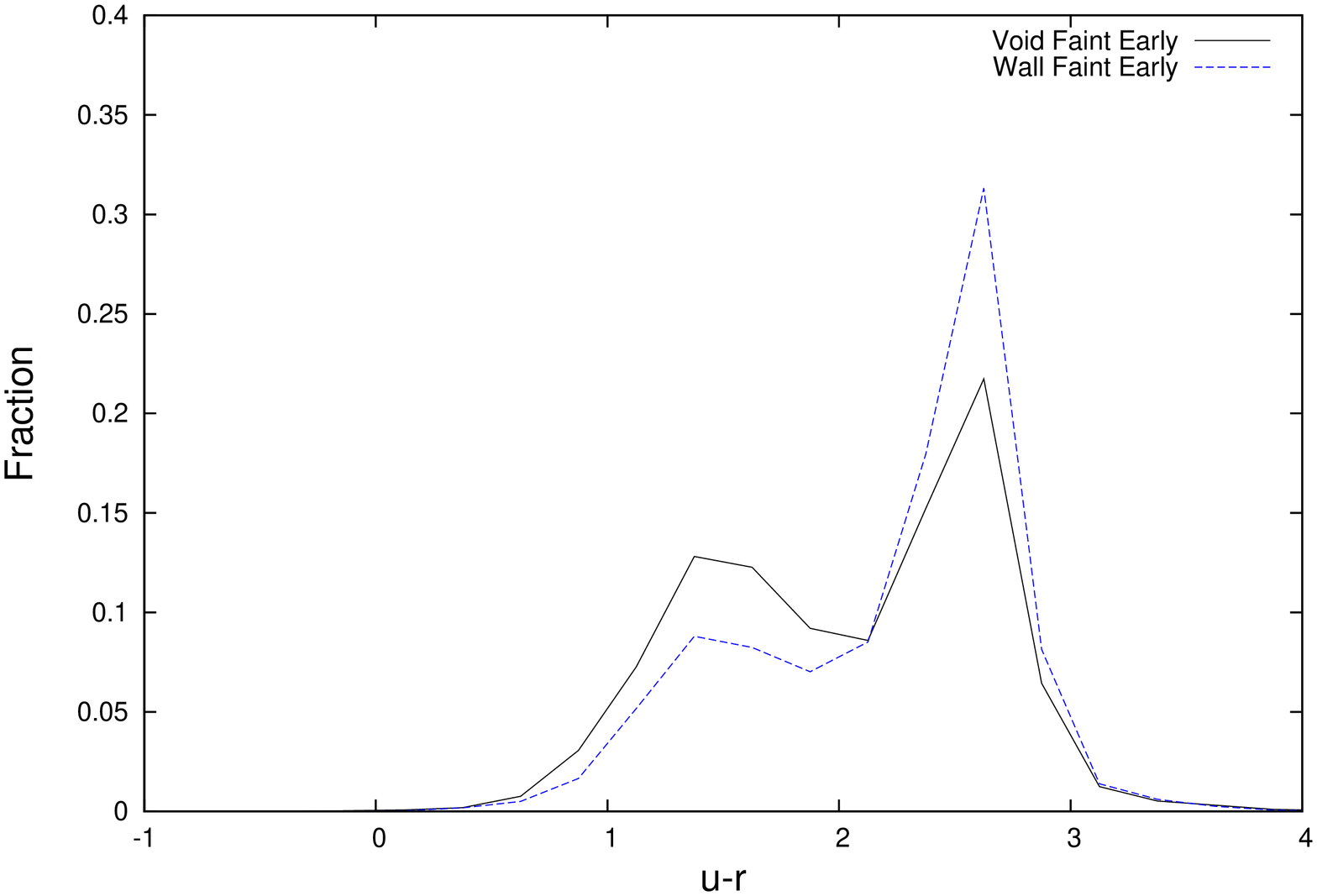}
&
\includegraphics[width=1.0in,angle=270]{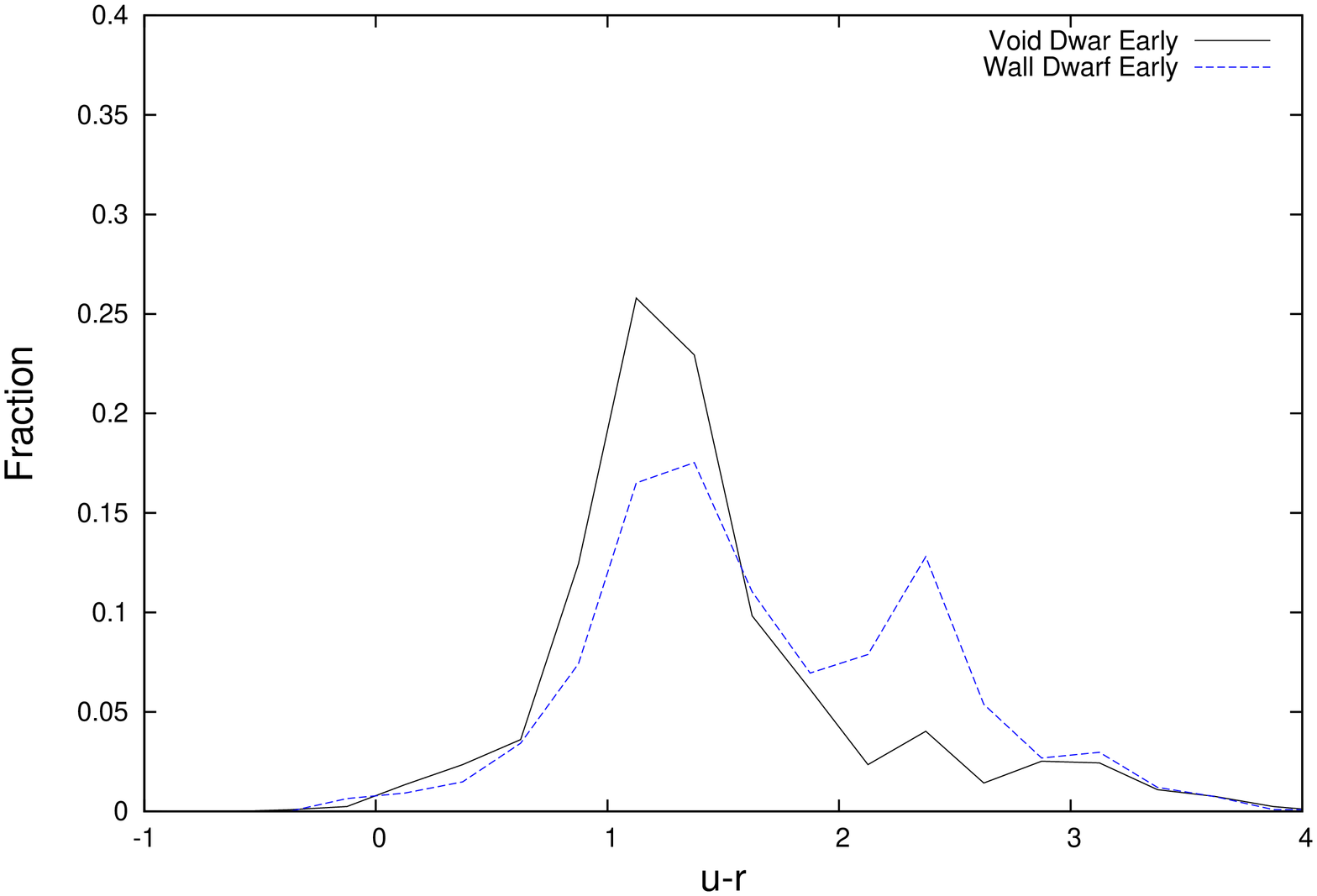} \\
\end{tabular}
\caption{The fraction of galaxies as a function of their $u - r$ colour 
 for early type (ICI $<$ 0.42) void galaxies (black, solid) and wall galaxies
 (blue, dashed) as a function of magnitude, as indicated in each figure.}
\label{fig:ureliptical}
\end{figure}

To investigate further the properties of void galaxies, we split each
of the samples (bright - dwarf) into sub samples of early or late
type galaxies, as measured by their inverse concentration index. We
split at ICI = 0.42, with galaxies with ICI less than 0.42 considered
early type and values greater than 0.42 are late type. In figure
~\ref{fig:urspiral} and ~\ref{fig:ureliptical} we show
the $u - r$ colour distribution for void galaxies (black solid) and wall
galaxies (blue, dashed) for galaxies with specific ICI as
described in the caption. In all cases, we see that there is a slight
shift to the blue for void galaxies as compared to wall galaxies. The differences in
colours by galaxy type are visually more striking for the faint galaxies than
the bright galaxies, although in the dwarf sample there are fewer
galaxies so the differences are less statistically significant. 

The late type galaxies shown in ~\ref{fig:urspiral} show a shift
towards the blue as the brightness of the sample decreases. This is
also true for the early type galaxies in ~\ref{fig:ureliptical}. It is also interesting 
that even though we cut by ICI, the bimodality of
the $u - r$ distribution is apparent again in the last two panels of figure
~\ref{fig:ureliptical}. It is not the case that
all early type galaxies are red. There exists a population of early type galaxies, as identified
by the ICI, that are faint and blue in colour. 

\begin{table*}
\centering 
\begin{tabular}{c c c c c c c c c }  
 & $u - r$ & sd   & $u - r$ & sd  &
ICI & sd  & ICI & sd \\ 
Sample & void & void & wall & wall & void & void & wall & wall \\
\hline 
All         & 2.043 & 0.002 & 2.162 & 0.002 & 0.4219 & 0.0002 & 0.4143 & 0.0002      \\
Bright    & 2.324 & 0.003 & 2.422 & 0.003 &  0.3980 & 0.0004 &  0.3889 & 0.0004 \\
Medium & 2.090 & 0.003 & 2.203 & 0.003 &  0.4223 & 0.0004 &  0.4121 & 0.0004 \\
Faint      & 1.786 & 0.003 & 1.917 & 0.003 &  0.4414 & 0.0004 &  0.4083 & 0.0004 \\
Dwarf & 1.422  &  0.008  & 1.589 & 0.009 & 0.464 & 0.001  &   0.466 & 0.001 \\
\hline 
\end{tabular} 
\caption{Mean colours and inverse concentration indices of void and wall
  galaxies. The columns sm representes the sample deviation, which is the standard deviation
  divided by the square root of the total number of galaxies in the sample. As the galaxy
sample is large, these values becomes small.} 
\label{table:photomean} 
\end{table*}

\begin{table*}
\centering
\begin{tabular}{cccccccccccccc}
Sample & $\mu_{blue}$ & $\mu_{red}$ & $\sigma_{blue}$ & $\sigma_{red}$
& norm$_{blue}$  & norm$_{red}$ & Sample & $\mu_{blue}$ & $\mu_{red}$ & $\sigma_{blue}$ & $\sigma_{red}$
& norm$_{blue}$  & norm$_{red}$ \\
\hline
Void All & 1.70 & 2.66 & 0.38 & 0.21 & 6.30 & 10.95 & Wall All & 1.76
& 2.69 & 0.40 & 0.20 & 7.00 & 9.33 \\
Void Bright & 2.04 & 2.77 & 0.42 & 0.15 & 6.55 & 10.26 & Wall Bright &
2.08 & 2.78 & 0.42 & 0.15 & 7.75 & 8.25 \\
Void Medium & 1.75 & 2.64 & 0.34 & 0.20 & 6.40 & 10.67 & Wall Medium &
1.80 & 2.67 & 0.36 & 0.19 & 7.40 & 8.71 \\
Void Faint & 1.55 & 2.51 & 0.31 & 0.24 & 5.25 & 16.80 & Wall Faint &
1.60 & 2.54 & 0.33 & 0.21 & 5.95 & 12.21 \\
Void Dwarf & 1.35 & 2.57 & 0.31 & 0.39 & 4.35 & 49.7 & Wall Dwarf &
1.39 & 2.25 & 0.31 & 0.28 & 5.25 & 16.80 \\
\hline 
\end{tabular} 
\caption{Gaussian fits to the various samples of void and wall
  galaxies. As the sample changes from bright to dwarf, the proportion
  of red galaxies decreases and the proportion of blue galaxies
  increases. The mean blue and red values of
  the void galaxies are always bluer than those of the wall
  galaxies, expect for the fit of the red wing of the dwarf galaxies,
  where there are very few galaxies to confine the fit. } 
\label{tab:gaussian} 
\end{table*}

\begin{table*} 
\centering 
\begin{tabular}{c c c c c c c c  }  
Sample& N & $u - r$  & sd & Galaxy Type & N & $u - r$ & sd \\
\hline 
All Late Void & 46209 & 1.770 & 0.002 & All Late Wall & 41793 & 1.835 & 0.002 \\
Bright Late Void & 11354 & 1.993 & 0.004 & Bright Late Wall & 9952 & 2.045 & 0.004 \\ 
Medium Late Void & 14816 & 1.824 & 0.003 & Medium Late Wall & 13046 & 1.880 & 0.004 \\
Faint Late Void & 16653 & 1.641 & 0.003 & Faint Late Wall & 15523 & 1.722 & 0.003 \\
Dwarf Late Void & 3386 & 1.417 & 0.008 & Dwarf Late Wall & 3272 & 1.554 & 0.009 \\ 
\hline 
\end{tabular} 
\caption{Mean colours of void and wall galaxies classified with inverse
concentration index $>$ 0.42. Sd represents the sample deviation, as
described in table ~\ref{table:photomean}} 
\label{table:urspiral} 
\end{table*}

\begin{table*}
\centering 
\begin{tabular}{c c c c c c c c  }  
Galaxy Type & N & $u - r$ & sd & Galaxy Type & N & $u - r$ & sd \\
\hline 
All Early Void & 42585 & 2.340 & 0.002 & All Early Wall & 46563 & 2.455 & 0.002 \\
Bright Early Void & 17921 & 2.534 & 0.003 & Bright Early Wall & 19488 & 2.614 & 0.003 \\ 
Medium Early Void & 13927 & 2.373 & 0.004 & Medium Early Wall & 15514 & 2.475 & 0.004 \\
Faint Early Void & 9546 & 2.039 & 0.006 & Faint Early Wall & 10482 & 2.207 & 0.005 \\
Dwarf Early Void & 1191 & 1.44 & 0.02 &Dwarf Early Wall & 1079 & 1.70 & 0.02 \\ 
\hline 
\end{tabular} 
\caption{Mean colours of void and wall galaxies classified with inverse
concentration index $<$ 0.42.  Sd represents the sample deviation, as
described in table ~\ref{table:photomean} } 
\label{table:urelliptical} 
\end{table*}

\subsection{Void Galaxy Properties as a function of Distance from
  Void Center}

One possible factor that could change the photometric properties of
the void galaxies is their location in the voids. We found above that
void galaxies are bluer than wall galaxies. Continuing this trend, perhaps the voids
closest to the center are bluer than those at the edge? In order to
investigate this possibility, we look at the histograms of distance from the void
center as a function of magnitude, colour and type. 

\begin{figure}
\centering
\begin{tabular}{c}
\includegraphics[width=2.0in,angle=270]{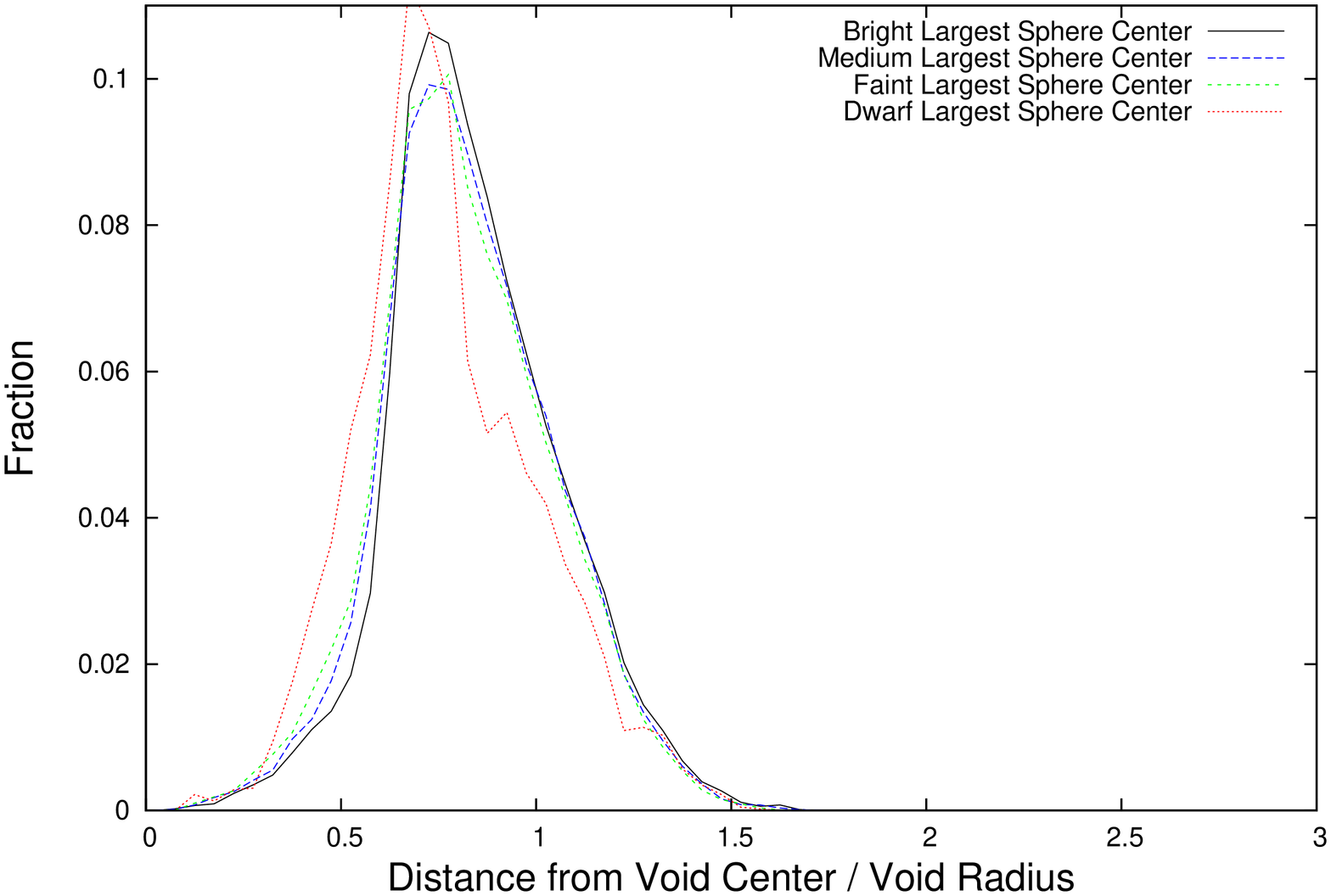}
\\
\includegraphics[width=2.0in,angle=270]{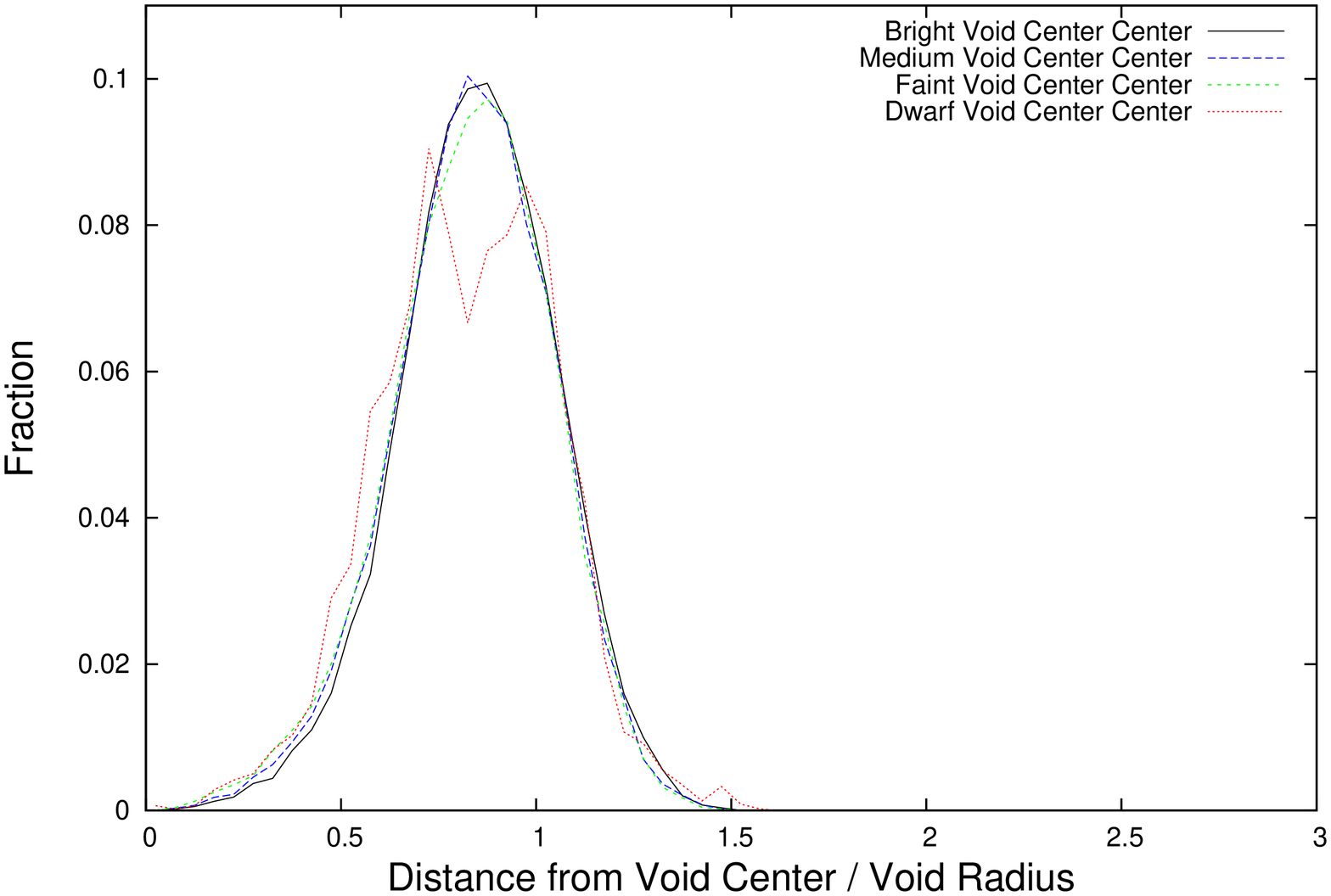}
\\
\end{tabular}
\caption{Histogram of the distance of a void galaxy from the center of its
 void, divided by the radius of that void. In the upper plot, the
 center of the void is defined as the center of the maximal sphere of
 the void. In the lower plot, the center of the sphere is
 defined as the center of mass as described in section
 ~\ref{sec:distancevoidcent}. The
different lines show the distribution for the different magnitude
samples. There is very little difference between the distributions,
except for the dwarf sample. }
\label{fig:distvoidcent}
\end{figure}

In figure ~\ref{fig:distvoidcent}, we show the histograms of the
distance from the void center for void galaxies with different
magnitudes. We determine the center of the void in two ways. In the 
upper figure we use the center of the maximal sphere as the center of the
void. In the lower figure we use the center of mass to define the
void center, as described earlier in section
~\ref{sec:distancevoidcent}. 

The peak of the distribution of void galaxy distances from their
center divided by the radius of the void is at ~87\%. About 20\% of
void galaxies have normalised distances (void center / void
radius) larger than 1 which might appear to indicate the the galaxies live
outside of the voids. This is not the case, because voids are not
constrained to be spherical although the effective radius of the void is
calculated assuming the void is a sphere. 

\begin{figure}
\centering
\begin{tabular}{c}
\includegraphics[width=2.0in,angle=270]{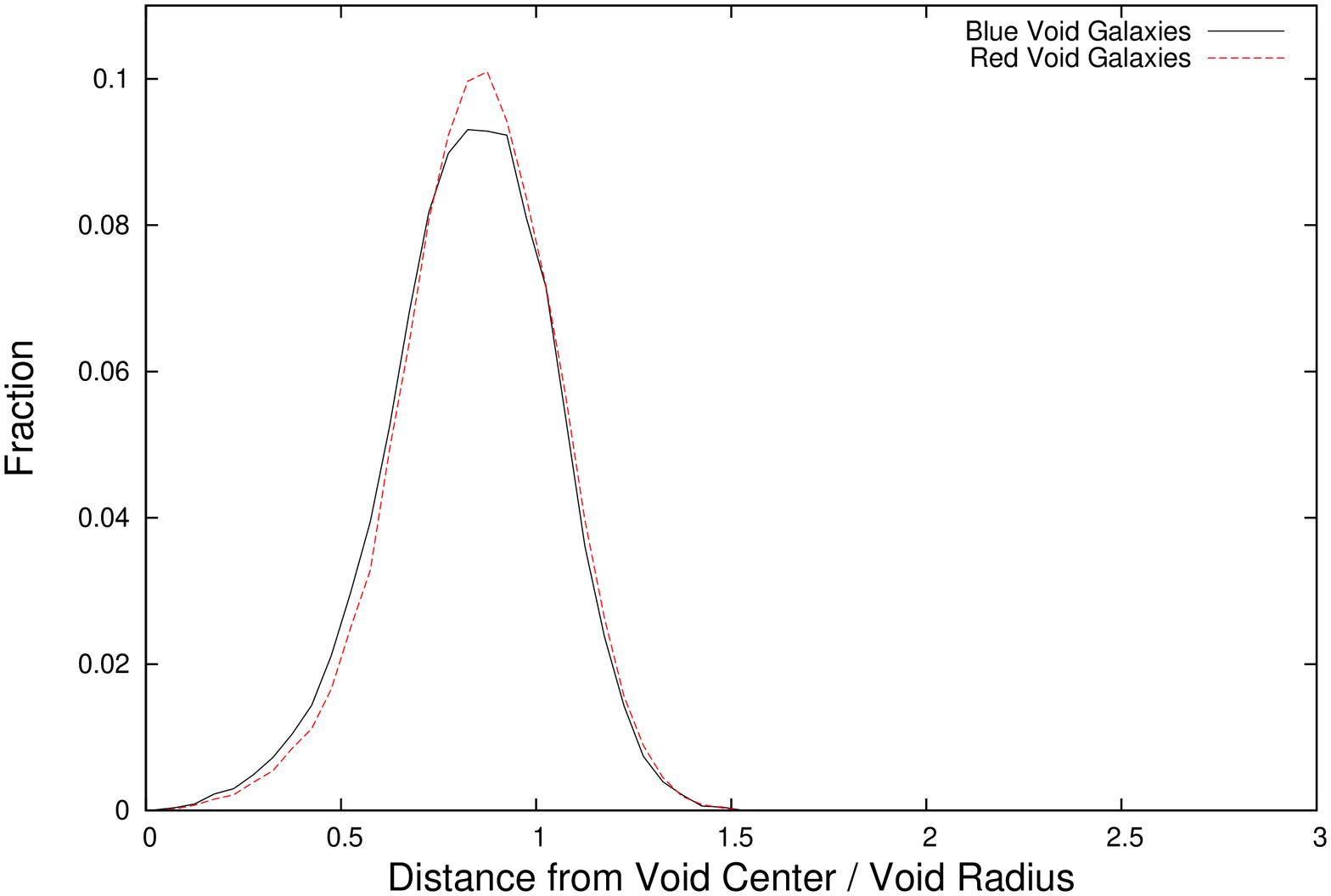}
\\
\includegraphics[width=2.0in,angle=270]{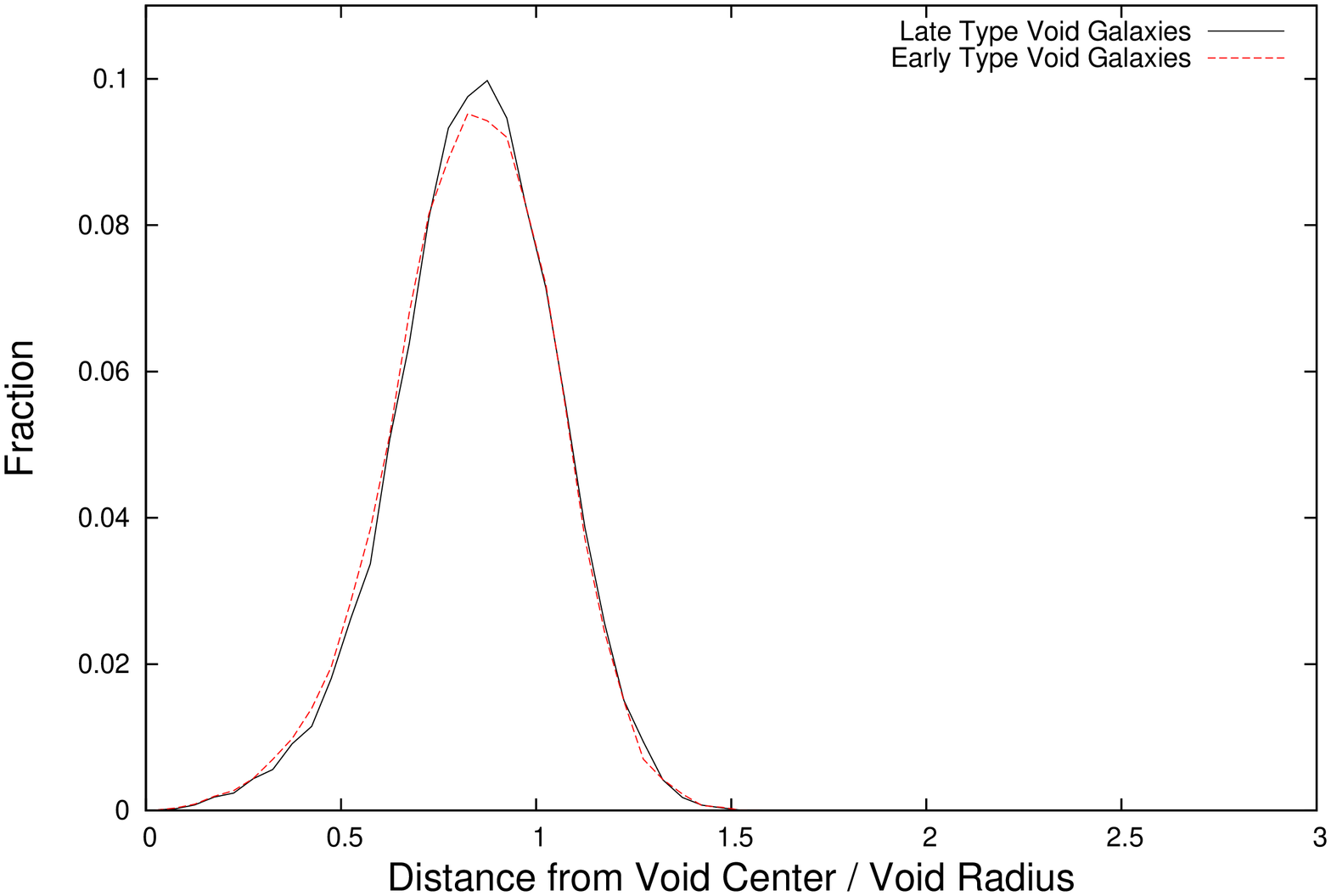}
\\
\end{tabular}
\caption{The upper plot shows the histogram of the distance of a void galaxy from the center of
 its void, divided by the radius of that void. There is little
 difference between the distribution of blue ($u - r <$ 2) and red ($u - r \ge$ 2) void
 galaxies. The lower plot shows the histogram of the distance of a void galaxy from the center of
 its void, divided by the radius of that void. Again there is little
 difference between the distribution of late (ICI $>$ 0.42) and early
 (ICI $\le$ 0.42) void galaxies}
\label{fig:distvtype}
\end{figure}

Figures ~\ref{fig:distvoidcent} and ~\ref{fig:distvtype} show that
magnitude, colour or type hardly alter
the shapes of the histograms. The histograms are very similar for
red/blue void galaxies and for late/early type void galaxies. The
bright, medium and faint samples also have similar distributions. 

\begin{figure}
\centering
\includegraphics[width=3.0in,angle=0]{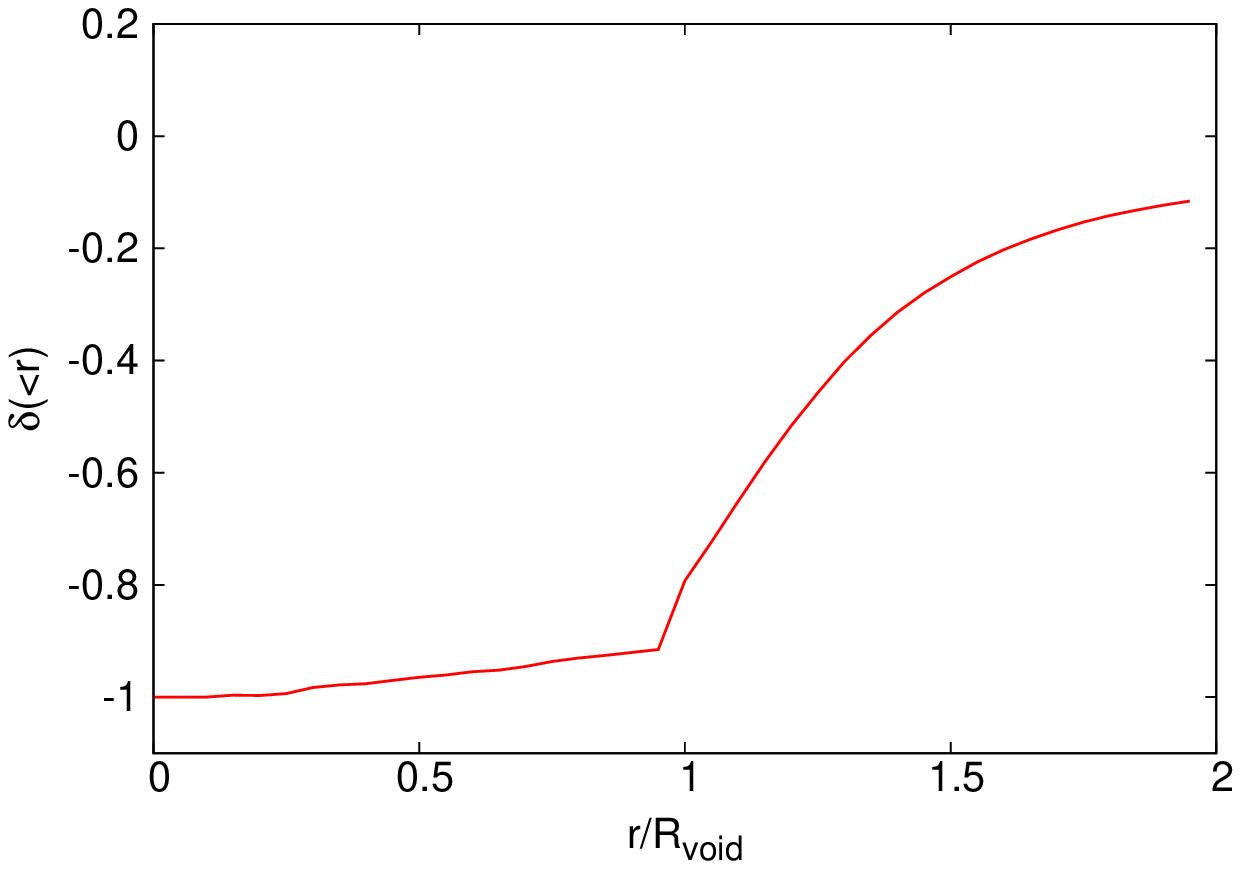}
\caption{The radial density profile of voids taken from ~\citep{Pan:2011}. The centers of voids are
very empty, with $\delta \rho / \rho$ close to -1, the lowest possible
value, until close to the radius. At the radius, the density of the void rises significantly.}
\label{fig:profile}
\end{figure}

At first it may seem surprising that the location of the void galaxy
{\it within} the void does not seem to impact its properties while the
properties of void galaxies as a group are different than those at
higher density: void galaxies are bluer and of slightly later type. 
However, as shown by ~\citet{Pan:2011}, and repeated in figure
~\ref{fig:profile}, voids are very empty, $\delta \rho / \rho < -0.9$,
and the density profile is essentially flat within the void. This
means that {\em within the void, galaxies reside in very similar density
environments.} 

The only histogram that does appear to differ is that of the dwarf
galaxies. When we calculate the center of the void
using the center of the maximal sphere, it appears that the dwarf galaxies possibly live closer
to the centers of the voids than the brighter galaxies. However, there
are fewer galaxies in this sample and when we use the center of mass
as the center of the void there is little difference between the distributions. 
This possible finding is, however, in agreement with high-resolution simulations by
~\citet{Kreckel:2011}, who find an excess of very faint, M$_r <
-14$, galaxies in the centers of voids.

\citet{Pan:thesis} found that Ly$\alpha$ systems also
prefer to reside at the center of cosmic voids. Of a sample of 119
absorbers from \citet{Danforth:2008} that overlap with the SDSS main sample, 87 lie in void
regions. Most of these absorbers are in close proximity to a nearby
void galaxy. Figure 6.3 of Pan (2011) shows that the absorbers are
not randomly distributed in the nearby Universe, but rather they have a preference to reside
towards the centers of the most underdense structures in the
Universe. The centers of voids are clearly an area that require more
investigation.

\section{Conclusions}
\label{sec:concs}

We examine a sample of 88,794 void galaxies that reside in void environments with
density contrast $\delta \rho / \rho < -0.9$. This is the largest
sample of void galaxies available at the moment and, with no redshift
survey of comparable sampling density in the pipeline, this sample will be the largest for a
while. 

The properties of void galaxies are compared for the first time to a
sample of galaxies at higher density that have the same magnitude
distribution as the void galaxies. This allows us to control for
differences due to luminosity alone. 

As we found in previous work, but with greater significance here, 
void galaxies are statistically bluer than galaxies found in higher 
density environments. This holds true for the whole sample of galaxies
as well as for samples of galaxies restricted to a narrow range of
magnitude. By fitting a double gaussian to the $u - r$ distribution of
galaxy colours in the voids and walls, we find that
there are more blue galaxies in the voids than in the walls in all
magnitude bins. As the galaxies decrease in magnitude, the blue
population grows in both void and wall environments.  

Void galaxies are also of slightly later type than wall galaxies, as measured by the
inverse concentration index. We split the void and wall galaxy samples
into late and early sub-samples and find that both late and early type void galaxies are
bluer than the comparison group of late and early type wall
galaxies. We find that faint, dwarf galaxies, classified as early by the ICI, are blue in colour. 

We have shown that the void environment affects that colours and types of the
galaxies. However, within the void there is little variation in the
properties of the galaxies as a function of distance from the center
of the void. Bright, medium and faint galaxies have a
very similar distribution of distances from the void center. Late and
early type galaxies within the voids have similar distance
distributions, as do galaxies split by colour. The lack of
difference is probably because the density profile of a void is very
flat in the center. Most of the galaxies within the voids live in
regions with the same density contrast. The only exception is for 
the faintest void galaxies, where there is a slight indication that
they are found closer to the centers of the voids.

\section*{Acknowledgments}

Funding for the creation and distribution of the SDSS Archive has been
provided by the Alfred P. Sloan Foundation, the Participating
Institutions, the National Aeronautics and Space Administration, the
National Science Foundation, the U.S. Department of Energy, the
Japanese Monbukagakusho, and the Max Planck Society. The SDSS Web site is http://www.sdss.org/.

The SDSS is managed by the Astrophysical Research Consortium (ARC) for
the Participating Institutions. The Participating Institutions are
The University of Chicago, Fermilab, the Institute for Advanced Study,
the Japan Participation Group, The Johns Hopkins University, Los
Alamos National Laboratory, the Max-Planck-Institute for Astronomy
(MPIA), the Max-Planck-Institute for Astrophysics (MPA), New Mexico
State University, University of Pittsburgh, Princeton University, the
United States Naval Observatory, and the University of Washington.

FH thanks the Engineering and Research Departments at PUCE for giving
me the opportunity to carry out this research.

\end{document}